\documentclass[10pt,journal]{IEEEtran}
\usepackage{balance}
\usepackage[ruled]{algorithm2e}
\usepackage{amssymb}
\usepackage{mathbbol}
\usepackage{stmaryrd}
\usepackage{cite}
\usepackage{color}
\usepackage{multirow}
\usepackage{subfigure}
\usepackage{graphicx,times,amsmath}
\usepackage{epstopdf}
\usepackage{pifont}

\ifCLASSINFOpdf

\else

\fi

\begin{document}

	\title{Enabling Low-Power OFDM for IoT by Exploiting Asymmetric Clock Rates}
		
		\author{Wei~Wang,~\IEEEmembership{Member,~IEEE,}
			Shiyue~He,
			Qian~Zhang,~\IEEEmembership{Fellow,~IEEE,}
			and~Tao~Jiang,~\IEEEmembership{Fellow,~IEEE}% <-this % stops a space

			\IEEEcompsocitemizethanks
			{
				\IEEEcompsocthanksitem Part of this work has been presented at IEEE INFOCOM \cite{wang2018wi}.
				\IEEEcompsocthanksitem This work was supported in part by the National Key R\&D Program of China under Grants 2017YFE0121500 and 2019YFB180003400, the Young Elite Scientists Sponsorship Program by CAST under Grant 2018QNRC001, the National Science Foundation of China with Grants 61729101 and 91738202, the RGC under Contract CERG 16203719 and 16204418, and the Guangdong Natural Science Foundation No. 2017A030312008.
				\IEEEcompsocthanksitem W. Wang, S. He, and T. Jiang are with the School of Electronic Information and Communications, Huazhong University of Science and Technology. E-mail: \{weiwangw, shiyue\_he, taojiang\}@hust.edu.cn.
				\IEEEcompsocthanksitem Q. Zhang is with the Department of Computer Science and Engineering, Hong Kong University of Science and Technology. E-mail: {qianzh}@cse.ust.hk.\protect\\
				}
			}
\maketitle
	%\sloppy
	
\begin{abstract} %2.27
The conventional high-speed Wi-Fi has recently become a contender for low-power Internet-of-Things (IoT) communications. OFDM continues its adoption in the new IoT Wi-Fi standard due to its spectrum efficiency that can support the demand of massive IoT connectivity. While the IoT Wi-Fi standard offers many new features to improve power and spectrum efficiency, the basic physical layer (PHY) structure of transceiver design still conforms to its conventional design rationale where access points (AP) and clients employ the same OFDM PHY. In this paper, we argue that current Wi-Fi PHY design does not take full advantage of the inherent asymmetry between AP and IoT. To fill the gap, we propose an asymmetric design where IoT devices transmit uplink packets using the lowest power while pushing all the decoding burdens to the AP side. Such a design utilizes the sufficient power and computational resources at AP to trade for the transmission (TX) power of IoT devices. The core technique enabling this asymmetric design is that the AP takes full power of its high clock rate to boost the decoding ability. 
We provide an implementation of our design and show that it can reduce up to 88\% of the IoT’s TX power when the AP sets $8\times$ clock rate. 

\end{abstract}

\section{Introduction}\label{sec:intro} %1.5 pp
%refer to UMASS backscatter or other assymetric design

%1. traditional wifi transceiver deisgns assume the same clock rate for both sides. we argue this restriction reaps no benefits for IoT networks.

%para1: we are entering the era of IoT
We are entering the post-PC era where an ever-larger variety of smart Internet-of-things (IoT) devices including wearables and smart sensors are increasing the demands for low power communication technologies. Wi-Fi has recently become a contender for this regime due to its compatibility with IP networks and wide deployments of hotspots. Growing numbers of wearables in the market have enabled Wi-Fi connection to the Internet without relying on gateways. Many commercial smartwatches including Apple Watch have already equipped with built-in Wi-Fi chipsets. Google's latest Android Wear 2.0 release~\cite{androidwear} allows Android smartwatches to directly connect to Wi-Fi networks.

With the proliferation of sensor-equipped smart objects in streets, homes, and offices, IoT connectivity is envisioned to be massive~\cite{cisco,wang2016less}. Thus, Wi-Fi for IoT devices should be not only low power but also spectrum efficient. As a mature and spectrum efficient multiplexing access technology widely adopted in the latest Wi-Fi standards, OFDM continues its adoption in the new Wi-Fi standard for IoT, i.e., IEEE 802.11ah~\cite{wifihalow}, which reuses OFDM frame format that conforms to IEEE 802.11ac to ensure spectrum efficiency.

\begin{figure}[t]
	\center
	\includegraphics[width=0.48\textwidth]{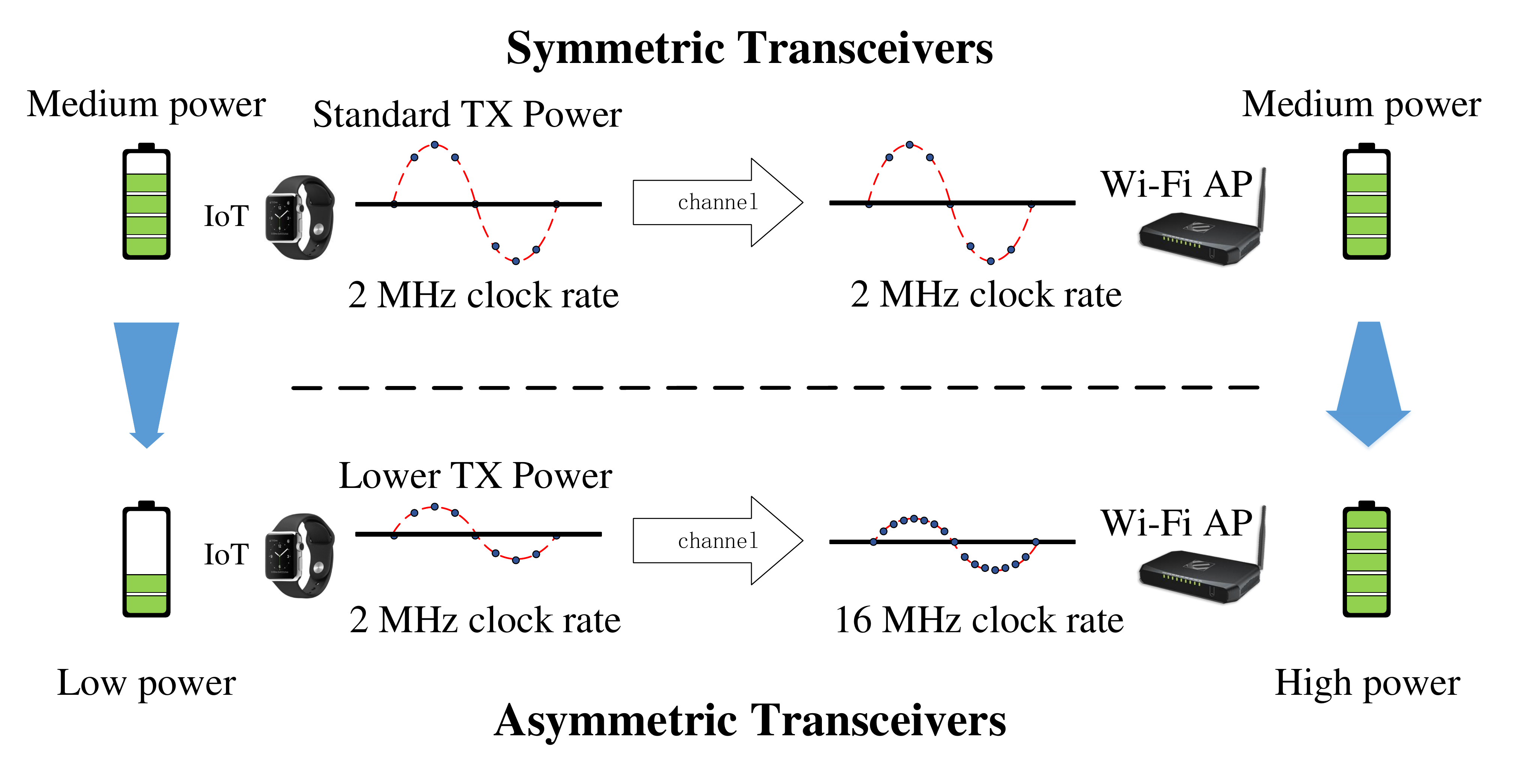}\vspace{0.3cm}
	\caption{The concept of asymmetric transceivers. A IoT device lowers its TX power and pushes all decoding and energy burdens to the AP side. As the compensation for decreased TX power, the AP takes full advantage of its high clock rate.}\label{fig:rationale}
\end{figure}

A number of recent research efforts have been devoted to reducing the power consumption of OFDM-based Wi-Fi communications. Downclocking receivers' radios is proposed to reduce the power consumption while receiving packets~\cite{wang2017sampleless,mobicom14enfold} or idle listening~\cite{mobicom11emili,wang2017wideband}. Efficient sleep modes~\cite{psm,khorov2015survey} are proposed to reduce the power consumption of conventional IEEE 802.11 nodes by allowing them to enter extremely low power state during idle listening. Despite these growing attempts and extensive efforts, most of them have focused on energy-efficient data reception. A variety of IoT applications require data-rich sensors such as cameras~\cite{arlo} and microphones to frequently upload sampled data to servers, which incurs a large amount of energy consumption for uplink transmission.

In this paper, we argue that the fundamental hurdle for energy efficient uplink transmission lies in the transceiver design. Existing Wi-Fi transceivers are originally designed for symmetric nodes with equal hardware capabilities and power constraints. This assumption no longer stands for IoT applications, where a Wi-Fi AP is much more powerful and almost energy-unconstrained compared to IoT devices. This evokes a similar picture in cellular networks, where base stations are equipped with 100$\times$ sensitive RF front-end and have 20~dB more transmission (TX) power than mobile clients. Analogous to the hardware asymmetry in cellular networks, APs are equipped with Wi-Fi chipsets that can support up to 160~MHz bandwidth in IEEE 11ac/11ax, while IoT devices normally support only 1-2MHz bandwidth~\cite{wifihalow}. Thus, there is significant potential to exploit asymmetric PHY configurations to enable low power uplink transmission without undermining decoding performance.

\begin{figure*}[t]
	\center
	\includegraphics[width=7in]{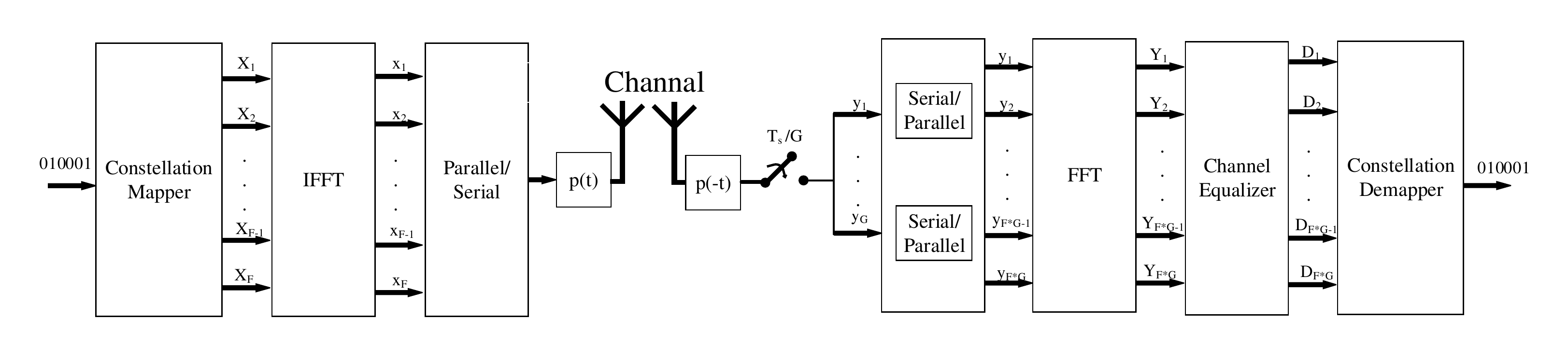}\vspace{0.3cm}
	\caption{OFDM schematic under overclocking.} \label{fig:ofdm_schematic} 
\end{figure*}

Our line of attack starts from the rationale that we trade the computing and energy resources of APs for the TX power of IoT devices in an \textit{asymmetric} fashion, that is, we allow IoT devices to transmit uplink packets using the lowest power while pushing all the decoding burdens to the AP side, as illustrated in Fig.~\ref{fig:rationale}. In particular, the AP takes full advantage of its clock rate that are tens of times higher than those of IoT's to decode packets with low signal-to-noise ratio (SNR). IoT devices reap benefits from such an asymmetric design by configuring TX power lower than the minimum power required for decoding. On the other hand, the AP with enough computing capabilities and constant power source can afford the extra cost induced by overclocking. Such an asymmetric design trades resources that are cheap to AP for power consumptions that are expensive to IoTs. Such a design completely conforms to Wi-Fi protocols, and thus can be readily integrated with existing standards. The only changes are standalone update of computational logics and RF settings at APs.

A key challenge in realizing the asymmetric design is how to effectively leverage the inherent overclocking potential in APs to decode conventionally undecodable packets with minimal modification to the conventional reception pipeline. Our fundamental insight is that APs yield correlated signals by exploiting the \textit{time shift} effect between overclocked samples. Specifically, when the AP sets its analog-to-digital converter (ADC) clock at a much higher rate than IoT devices, it yields multiple interpolated samples from one transmitted sample. These interpolated samples can be considered as time-shifted versions of the transmitted sample. When transformed into the frequency domain, time-shifted samples result in different phase rotations at different subcarriers. Thus, we can reuse the existing module to process the signal in the time domain. After transferring the signal to the frequency domain, we can leverage this phase rotation effect to reuse the packet decoding module. Another challenge stems from the lack of knowledge in modeling the noise distributions between these redundant samples. Instead of blindly reusing conventional decoders, we turn to a data-driven approach. In particular, we build a noise map from preambles without making distribution assumptions. Then, we combine redundant samples by employing a maximum likelihood (ML) decoder based on the joint probability of these samples.

We implement the above asymmetric design, referred to as \texttt{T-Fi}, on the GNURadio/USRP platform. Evaluation results validate \texttt{T-Fi} in reliably receiving and decoding Wi-Fi packets at low SNR across a wide range of scenarios. Furthermore, \texttt{T-Fi} can reduce the IoT's TX power when the AP is overclocked by a factor of up to 8, which is still lower than the rate used for IEEE 802.11a/g/n.

The contributions of this paper are summarized below:
\begin{itemize}
	\item We provide a thoughtful study towards enabling low-power OFDM transmissions for IoT Wi-Fi standards. Our solution can be seamlessly integrated into existing Wi-Fi standards without modifying the legacy frames or protocols. 
	\item We explore the fundamental structure of overclocked reception in OFDM, and propose a reception pipeline to decode legacy packets at lower SNRs than the conventional transceivers. The key underlying technique is a new decoding algorithm that exploits the time shift effects in oversampled signals.
	\item We build a full prototype of \texttt{T-Fi} and quantify the merits of our design in a wide range of scenarios.
\end{itemize}

\begin{figure}[t]
	\center
	\includegraphics[width=3.5in]{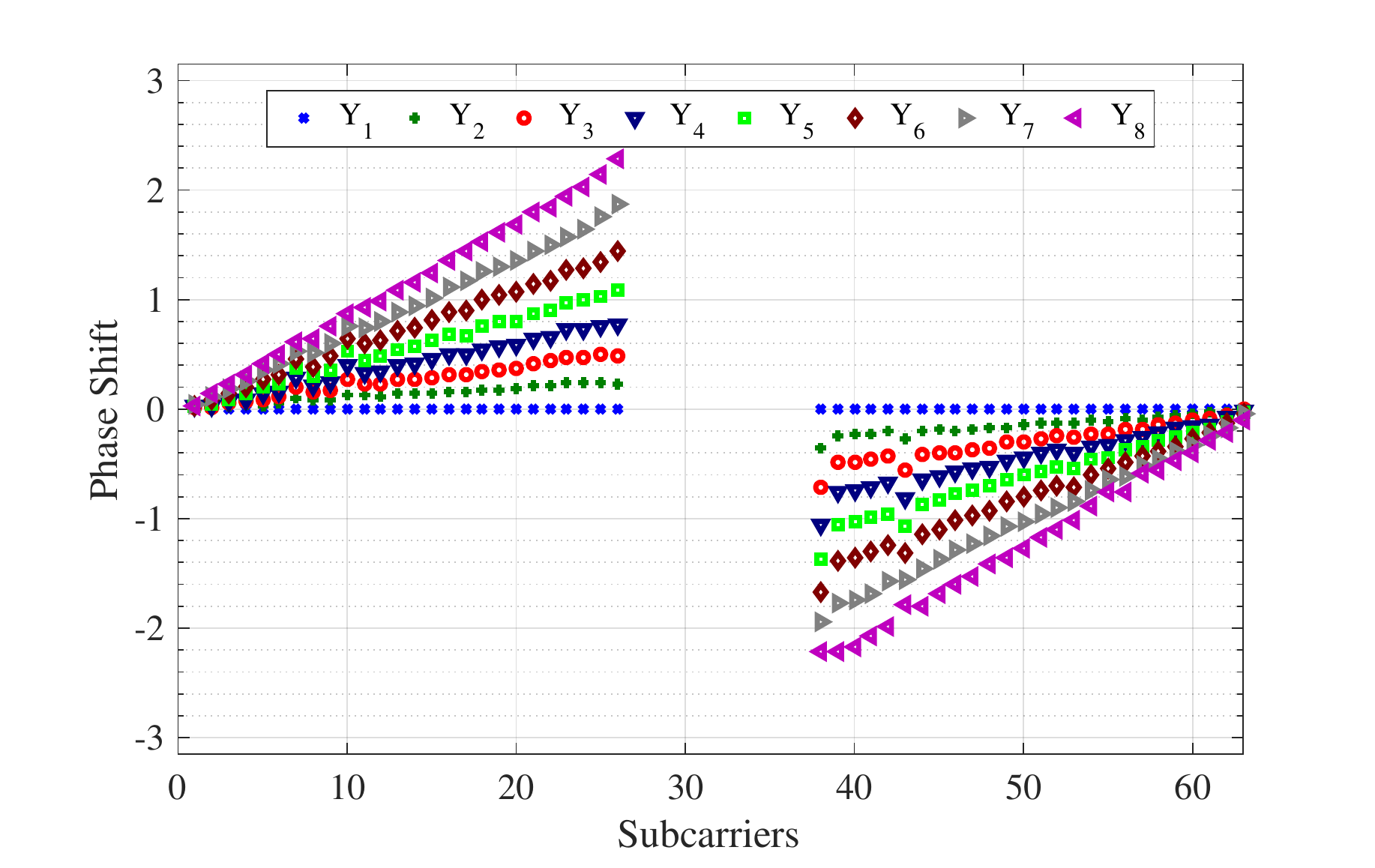}\vspace{0.3cm}
	\caption{Phase shift induced by oversampling a real Wi-Fi packet. Phases are normalized according to the phases of $Y_1$. Subcarriers 0-25, 38-63 carrying payload data are compared.} \label{fig:phaseshift}
\end{figure}

The remainder of the paper is structured as follows. We begin in Section~\ref{sec:motivation} with the design motivation. Section~\ref{sec:overclock} analyzes the effect of overclocking in OFDM, which is the underpinnings of our design. Section~\ref{sec:design} elaborates the detailed reception pipeline of our design. System implementation and experimental evaluation are introduced in Section~\ref{sec:implementation} and Section~\ref{sec:evaluation}, respectively. Section~\ref{sec:relatedwork} gives a brief survey of related work, followed by the discussion in Section~\ref{sec:discussion}. Section~\ref{sec:conclusion} concludes this work.

\vspace{0.3cm}

\begin{figure}[t]
	\center
	\includegraphics[width=3.5in]{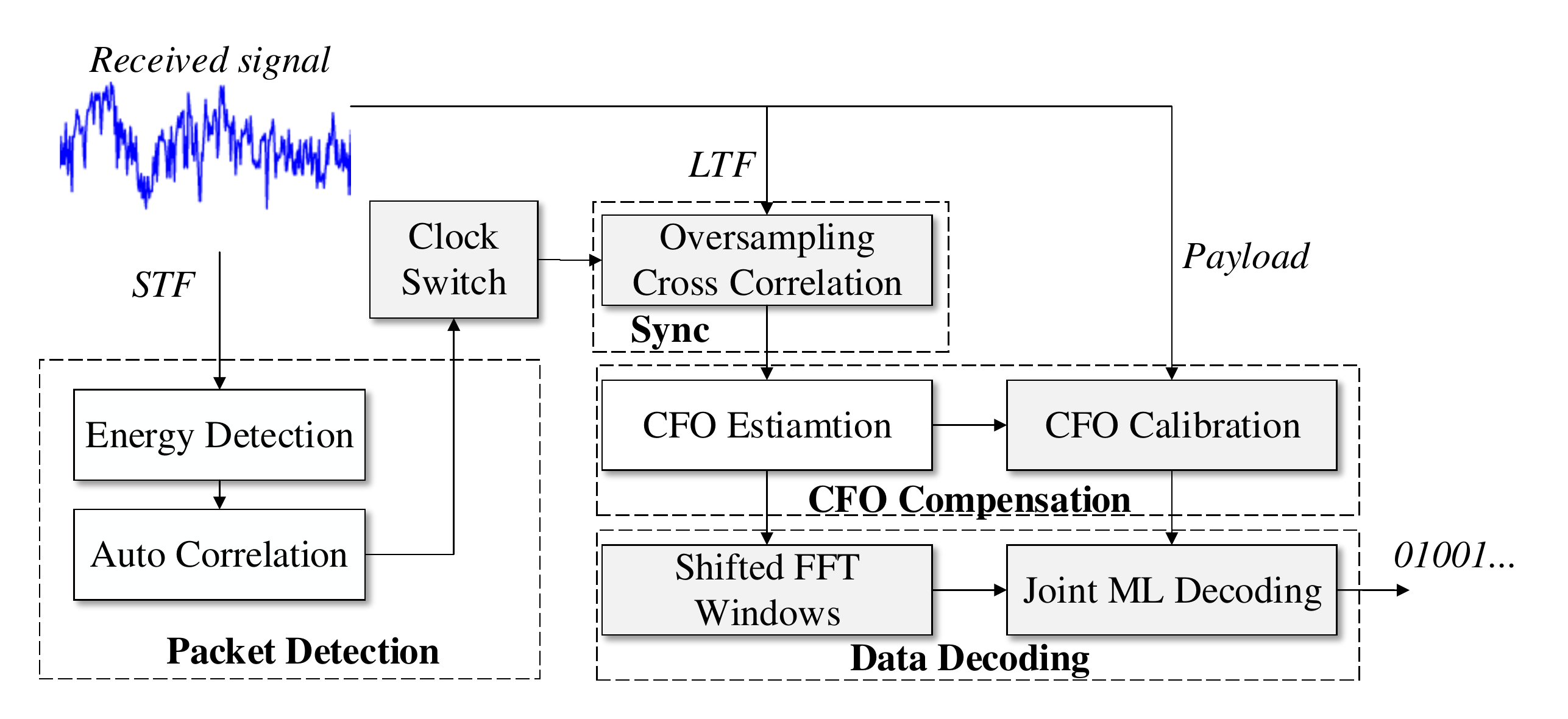}\vspace{0.3cm}
	\caption{System architecture of \texttt{T-Fi}.} \label{fig:architecture}\vspace{0.3cm}
\end{figure}

\section{Motivation of Transceiver Asymmetry}\label{sec:motivation}%1 pp

In this section, we ask and answer questions to explore practical and suitable transceiver architectures. Through this exploration, we hope to convince readers that the \textbf{asymmetric} design is a practical design in that it elegantly fits the architecture of IoT Wi-Fi while imposing affordable costs.

\subsection{IoT Scenarios and Requirements}

It is envisioned that the number of smart devices that need to wirelessly share data or access the Internet is growing exponentially. To embrace the coming wave of \textit{massive numbers} of wearables, driverless cars, smart sensors, IEEE 802.11ah was announced in 2016 to tailor Wi-Fi protocols specially for these IoT devices. The target envisioned in this IoT Wi-Fi standard is to challenge Bluetooth and cellular networks by achieving the best of both worlds: to enable \textit{long-distance} communications using relatively \textit{low amounts of power}. Specifically, IEEE 802.11ah defines the following requirements: (i) more than a fourfold number of devices supported by one AP compared to legacy Wi-Fi, (ii) doubled transmission range compared to legacy Wi-Fi links, and (iii) low energy consumption with guaranteed data rates of at least 100~Kb/s.

To meet the above requirements, IEEE 802.11ah introduces a bundle of new features in PHY/MAC. Particularly, the new standard reuses the formerly-adopted OFDM waveform consisting of 32/64 subcarriers while tailoring it to a narrowband channel of 1~MHz or 2~MHz in 900~MHz unlicensed spectrum. Comparing with other low power protocols~\cite{ratelesstodistanceless, ratelesstohopless} with low spectrum efficiency, OFDM is used to maintain spectrum efficiency for massive connectivity, and narrowband transmission at lower frequencies are used to facilitate low-power, long-range transmissions. In addition, a set of MAC mechanisms is designed to reduce the power consumption in channel access.

\subsection{Hardware Asymmetry}

Although the new standard specifies new PHY and MAC techniques and features to meet the requirements of IoTs, Wi-Fi transceivers still work in a conventional fashion: today's Wi-Fi radios work in a symmetric manner in that the transmitter and receiver are configured with the same clock rate. This symmetric transceiver design has worked well over the past decade as both sides are expected to be (more or less) equally powerful in conventional scenarios. While this assumption no longer stands in IoT scenarios where the disparity between the hardware capabilities of APs and IoTs makes such a design very inefficient. It is either an overwhelming burden for IoTs to match the radio configurations of APs, or makes APs largely under-utilize their hardware capabilities to shoehorn the proper settings of IoTs' radios.

The above situation evokes a similar picture in cellular networks, where base stations are much more powerful than mobile stations and are designed to support thousands of hardware-constrained mobile devices. Cellular networks equip base stations with higher TX power, RF sensitivity, and large form-factor antennas to compensate the hardware gap. The wisdom of cellular networks motivates us to design an asymmetric PHY that fully reap the benefits of the hardware gap between APs and IoTs. 

Our target is to explore an asymmetric transceiver design that can better fit the IoT scenarios and be seamlessly integrated into existing and future OFDM-based Wi-Fi protocols. The fundamental insight is that APs have already equipped with high clock-rate radios to support high-speed Wi-Fi such as IEEE 802.11ac/ax, and the clock rates are tens of times to those of IoTs'. When APs use such overclocked rates to sample IoT's transmissions, a large amount of redundancy between these samples can be used to boost the performance of packet reception, which in turn relaxes TX power requirements imposed on IoTs.

\vspace{0.3cm}

\section{Exploiting Overclocking Opportunities}\label{sec:overclock}% 1 pp

As the core of the asymmetric transceiver design, we exploit the redundancy in OFDM signals sampled by overclocked radios. To this end, we discuss the opportunities in retaining the OFDM samples at an overclocked rate and utilizing them to improve symbol decoding performance. We start by analyzing the correlations among these samples.

In a standard OFDM system, as illustrated in Fig.~\ref{fig:ofdm_schematic}, one OFDM symbol contains a sequence of bits that are modulated into a set of lattice points $X[f]$ on orthogonal subcarriers in the frequency domain. 
The OFDM symbol is transmitted and propagates a wireless channel with frequency response $H(f)$. In the presence of a time-dispersive channel and additive noise, the received continuous time-domain baseband signal $y(t)$ can be expressed as
\begin{equation}
	y(t) = \frac{1}{F} \sum_{f=0}^{F-1}{X[f] H(f) e^{j \frac{2\pi f t}{T}} } + w(t), 0 < t < T, \notag
\end{equation}
where $T$ and $w(t)$ are the symbol duration and complex Gaussian noise, respectively. Note that we omit cyclic prefix and the matched filter for simple illustration.

Now we consider a concrete example in which $F=64$ subcarriers are used to convey information. When using the same clock rate to sample received signal $y(t)$, the sampling instances are at $t={n \over 64} T$. Thus, the receiver yields
\begin{equation}
	y_1[n] = {1 \over 64} \sum_{f=0}^{63} X[f] H(f) e^{j {2\pi f n \over 64}  } + w_1(n).\notag
\end{equation}
When the receiver doubles the clock rate, the number of samples in each FFT segment is 128. 
As such, the time domain sample sequence is expressed as
\begin{align}
	y[n] =& {1 \over 64} \sum_{f=0}^{63} X[f] H(f) e^{j {2\pi f n \over 128}  } + w(n), n = 0,...,127 \notag\\
	= & {1 \over 64} \sum_{f=0}^{63} \left( X[f] H(f) e^{j {2\pi f  \over 128} 2n } + X[f] H(f) e^{j {2\pi f  \over 128} (2n+1) }  \right) \notag\\
	&+ w(2n) + w(2n+1), n = 0,...,63 \notag\\
	= & y_1[n] + y_2[n] + w_1(n) + w_2(n), n = 0,...,63, \notag
\end{align}
where $y_2[n] = {1 \over 64} \sum_{f=0}^{63} X[f] H(f) e^{j {2\pi f  \over 64} (n+1/2)}$ and $w_2(n)$ is the new noise sequence in oversampled signals. We see that $y_2[n]$ is a time-shifted version (delayed by a half sample) of $y_1[n]$. If the frequency response of $y_1[n]$ is defined as $Y_1[l]$, the frequency response of $y_2[n]$ is expressed as
\begin{align}
	Y_2[l] &= \sum_{n=0}^{63} y_2[n] e^{-{j 2\pi n l \over 64}}, l=0,...,63 \notag \\
	&= X[l] H(l) e^{-{j \pi l \over 64}} +  W_2(l).\notag
\end{align}
Ignoring the noises, $Y_2[l]$ is a phase-shifted version of $Y_1[l]$, and the amount of phase rotation is linear to the subcarrier frequency $l$. 

Hence, doubling the receiver's clock rate yields two receptions, which transforms the system into a single-input-multiple-output (SIMO) system. When we employ $G$-fold clock rate at the receiver, we can obtain $G$ phase-shifted versions of $Y_1[f]$ described as
\begin{equation}\label{e:yg}
	Y_g[f]	= X[f] H(f) e^{-{j 2 \pi gf \over 64G}} +  W_g(f), g=0,...,G-1.
\end{equation}
Fig.~\ref{fig:phaseshift} shows the phase shifts across all subcarriers in a real Wi-Fi packet when received at eight-fold clock rate using USRP testbeds. The predictable phase shifts in $Y_g[f]$ can be easily compensated to obtain $G$ copies of the transmitted signal. We can combine these copies to improve the received signal strength while amortizing noise, and thereby enhance the decoding capability under low SNR conditions.

We can analyze the power gain of oversampling from Eq.~\ref{e:yg}. Similar to the signals received by multiple antennas in SIMO system, we combine all signals with weight $\alpha_g$ after compensating the phase shift. The total received SNR at $G$-fold clock rate is
\begin{align}\label{e:snr1}
	\text{SNR}_G = \frac{(\sum_{g=0}^{G-1}\alpha_g X[f]H(f))^2}{N_0 (\sum_{g=0}^{G-1}\alpha_g^2)},
\end{align}
where $N_0$ is the noise power. 
Since the interpolated signals come from the same transmitter and pass through a channel with the same fading, there is no difference in the received power of each copy.
We let $\alpha_0 = \ldots = \alpha_{G-1} = \sqrt{E_0 / N_0}$, where $E_0$ is the power of the received signal at normal clock rate. Then, Eq.~\ref{e:snr1} is expressed as
\begin{align} \label{e:snr2}
	\text{SNR}_G 
	&= \frac{(\sum_{g=0}^{G-1} \sqrt{\frac{E_0}{N_0}} X[f]H(f))^2}{N_0 (\sum_{g=0}^{G-1}\frac{E_0}{N_0})} \notag\\
	&= \frac{(G \frac{E_0}{\sqrt{N_0}} )^2}{G E_0} 
	= G \frac{E_0}{N_0} = G\text{SNR}_1,
\end{align}
where $\text{SNR}_1$ denotes the SNR when the receiver uses the normal clock rate.
We can find that the power gain we are expecting to get is similar to the array gain in the MIMO system. In particular, we can get 3~dB power gain whenever we double the clock rate.

\vspace{0.3cm}

\section{Asymmetric Transceiver Design}
\label{sec:design}%3.5-4.5 pp

In this section, we describe our design of \texttt{T-Fi}, an asymmetric transceiver architecture for IEEE 802.11ah protocol. \texttt{T-Fi} can fully interoperate with standard IEEE 802.11ah devices, with no modifications to existing protocols. \texttt{T-Fi} leverages the high clock rate of APs to enable low power transmissions for IoT devices. This section elaborates the detailed reception pipeline design that embraces this design rationale.

\subsection{Overview}
An overview of \texttt{T-Fi} system architecture is shown in Fig.~\ref{fig:architecture}. \texttt{T-Fi} leverages the AP's high clock rate and processing capability to grant lower TX power for IoT devices. The reception pipeline conforms to the logic of a standard Wi-Fi receiver but is tailored to the asymmetric design to make the most of the redundancy induced by overclocking. An AP uses the normal clock rate for packet detection with the Short Training Field (STF) and then switches to the high clock rate mode after successfully detecting the packet. Under the overclocking setting, the AP yields redundant Long Training Field (LTF) and payload samples, and exploits the correlations among these samples to boost the performance of synchronization and carrier frequency offset (CFO) estimation under low SNR conditions. Then, the synchronization algorithm is performed on the LTF, while the CFO estimation is performed on the LTF and then calibrated in the following data symbols in the payload. Multiple copies of overclocked samples are fed to the data-driven ML decoder. We will introduce the design of each module in the rest of this section.

\begin{figure}[t]
	\center
	\includegraphics[width=3.5in]{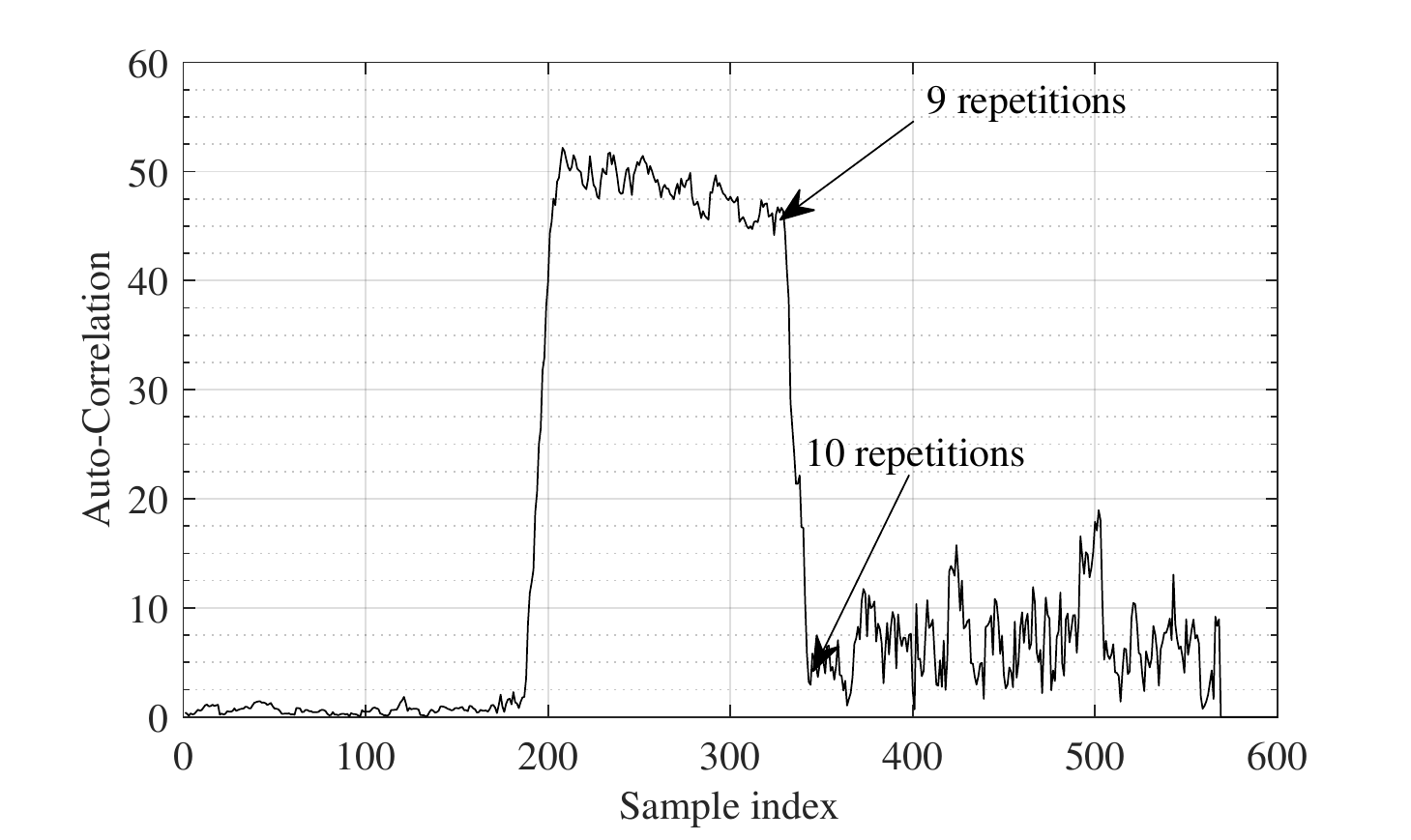}\vspace{0.3cm}
	\caption{Auto correlation response.} \label{fig:autocorr} %\vspace{0.1cm}
\end{figure}

\subsection{Timing Synchronization and Clock Switch}

Timing synchronization aims to detect the presence of a Wi-Fi packet and then identify the start of OFDM symbols. It measures the energy of the sampled signal to detect the packet. If the energy is higher than a given threshold, it then performs the auto-correlation to check whether the incoming signal is a Wi-Fi frame. For a Wi-Fi packet, the auto-correlation algorithm spikes and forms a plateau. Our experiments show that even at very low SNR, the plateau is still clear to be separated from noise.

We set normal clock rate for idle listening and packet detection to avoid unnecessary energy consumption.
After packet detection, the receiver switches to the overclocking mode, which incurs latency for clock switch. As the receiver keeps the same frequency synthesizer and the center frequency of its analog circuit, the latency comes from the digital phase-locked loop (PLL) stabilization. Wi-Fi radios take merely several microseconds (e.g., 8~$\mu s$ in MAXIM2831~\cite{maxim}) to stabilize PLL. 
On the other hand, a preamble consists of a STF and a LTF. The legacy STF contains two OFDM symbols that are comprised of ten repetitions of a 16-sample sequence, and the LTF contains two identical 64-sample (80-sample including cyclic prefix) OFDM symbols.
As the state-of-the-art IoT Wi-Fi, i.e., IEEE 802.11ah, shrinks the conventional 20~MHz bandwidth to 2~MHz while retaining the same number of subcarriers, the duration of one OFDM symbol is extended to 40~$\mu s$. Both STF and LTF last 80~$\mu s$, which leaves enough time for clock switch. As illustrated in Fig.~\ref{fig:autocorr}, which is tested with $\text{SNR} = 10$ dB, we observe that the first nine repetitions are enough to produce a plateau for packet detection. Thus, we take the first nine repetitions in STF to perform auto correlation while leaving the last repetition of STF for clock switch. In our experiment results, we show that it does not need to extra packet loss under low SNR conditions.

\begin{figure}[t]
	\center
	\includegraphics[width=3.5in]{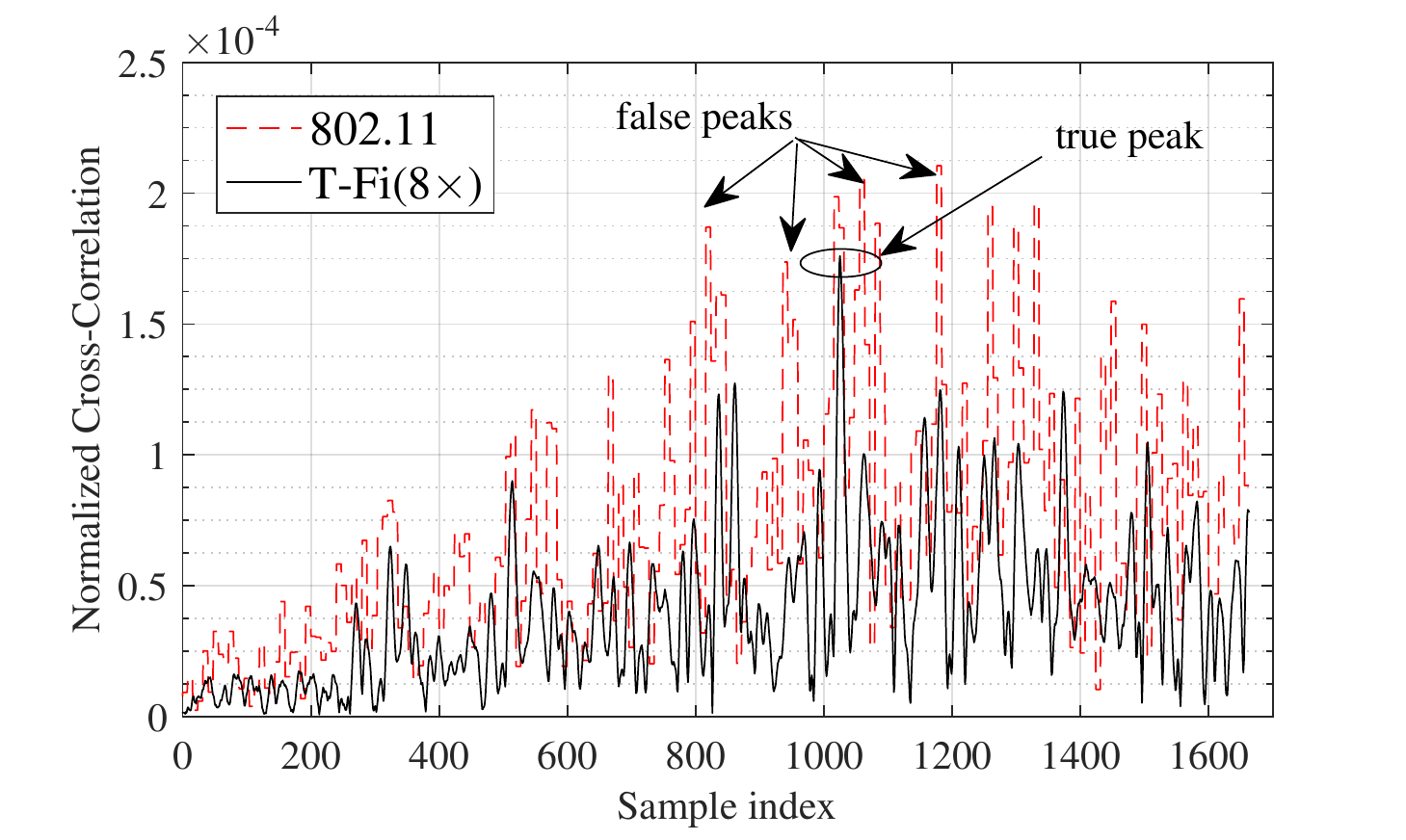}
	\caption{Cross correlation responses at various clock rates. We set up a USRP link to capture raw data samples pf a real Wi-Fi packet at 8$\times$ and 1$\times$ Nyquist rates, respectively.}\label{fig:sync_eg} 
\end{figure}

After successfully switching to the overclocked mode, the receiver precisely locates the boundary between OFDM symbols. While the IEEE 802.11 specifications do not mandate any specific algorithm, a typical synchronization algorithm is to use the cross-correlation property of the LTF. However, when the SNR is poor, cross correlation cannot cancel out noise, thereby resulting in multiple false peaks. We exploit correlations in oversampled LTF signals to overcome this predicament. In specific, when the LTF is sampled by an overclocked ADC, the correlation between oversampled data can be described by Eq.~\ref{e:yg}. Hence, the receiver can treat all the samples as a new LTF, which has stronger correlation properties. The cross correlation of the oversampled LTF can be represented as
\begin{align}
	c_{cross}[k] = \frac{\sum_{k=1}^{L}y[n+k]r^{*}[n]}{\sum_{k=1}^{L}\|r^{*}[n]\|^2},
\end{align}
where $r^{*}[n]$ is the new LTF with interpolated samples.
Fig.~\ref{fig:sync_eg} illustrates the merits of using the oversampled LTF for synchronization. Under low SNR conditions, the cross correlation result of a standard LTF produces multiple comparable peaks, which makes the receiver easily aligns to a wrong peak. The cross correlation result of an oversampled LTF produces a single highest peak corresponding to the full alignment of OFDM symbols.

\subsection{Frequency Offset Compensation}

CFO varies over time and must be estimated and compensated for each frame. In practice, the received baseband signal, instead of being centered at DC (0 Hz), is centered at a frequency offset $\Delta f$. Thus, the receiver yields
\begin{align}
	y_{\text{CFO}}(t) &= y(t)e^{\frac{j2\pi\Delta ft}{F_{s}}},\notag
\end{align}
where $F_s$ is the sampling frequency. CFO induces phase rotation over time that not only undermines the payload decoding but also affects the phase correlation among $Y_g[f]$. As our overclocking design relies on the phase correlation, CFO must be precisely estimated and calibrated.

Standard Wi-Fi receiver estimates CFO by comparing the phase rotation between the two identical OFDM symbols in LTF. Such an estimation is accurate enough for normal packet decoding, while the residual offset cannot be neglected when using the phase correlation among oversampled data. We make a fine-grained calibration in the following data symbols by exploiting the CFO effect among oversampled sequences $y_g[n]$. Let $\mathbf{y}_g = [y_g[0],...,y_g[N-1]]^\top$ denotes $g$th copy of the oversample signals, and the received data values in frequency domain $\mathbf{d} = [X[0]H(0),...,X[N-1]H(N-1)]^\top$. FFT is expressed in the form of a $N \times N$ matrix $\mathbf{F}$, where each entry $f_{ij}=e^{j (i-1)(j-1)}$. The effect of CFO on the first copy can be expressed as $\mathbf{P}=\text{diag}\left(1 \; e^{j {\Delta f\over F_{s}}} \cdot \cdot \cdot e^{j {\Delta f (N-1)\over F_{s}}}\right)$. The phase shifts due to overclocking can be described by $\mathbf{O}_g=\text{diag}\left(1 \; e^{-{j 2 \pi g \over NG}} \cdot \cdot \cdot  e^{-{j 2 \pi g(N-1) \over NG}} \right)$. Then, the $g$th copy of the oversample signals $\mathbf{y}_g$ can be expressed as
\begin{equation}
	\mathbf{y}_g = e^{\frac{j2\pi\Delta fg}{F_{s}G}} \mathbf{P} \mathbf{F} \mathbf{O}_g \mathbf{d} + \mathbf{W_g}.
\end{equation}
In the absence of noise, we obtain the following the relationship
\begin{align}\label{e:cfo}
	\mathbf{F}^{\text{H}} \mathbf{P}^{\text{H}} \mathbf{y}_g &= e^{-\frac{j2\pi\Delta fg}{F_{s}G}}  \mathbf{O}_g^{\text{H}} \mathbf{F}^{\text{H}} \mathbf{P}^{\text{H}} \mathbf{y_0},
\end{align}
where $(\cdot)^{\text{H}}$ is the conjugate transpose of a matrix. It is intuitive to find the unique $\Delta f$ that satisfying the above equation. In the presence of noises, we minimize the sum of distances between the left and the right hands of Eq.~\ref{e:cfo} for all $g>0$. We employ ML estimator to derive $\Delta f$.

\subsection{Decoding}\label{sec:decoding}% 1-2 pp
%- ML decoder
%- MRC decoder

After timing synchronization and frequency offset compensation, it is ineffective to apply the standard decoder in \texttt{T-Fi}. Under the overclocking setting, a receiver obtains multiple copies of each OFDM symbol with different delays. The effect of time shift on signal distortion may lead to an increase in the BER of the received signal~\cite{timesync}. 

In our experiments, we find that the signal distortion occurs in interpolated samples at the overclocking receiver. The signal distortion is caused by the time delay. At the standard receiver, such time-delay effect will not impact the decoding when the time synchronization is accurate. However, the interpolated samples between the $n$th OFDM symbol and the next are affected by two symbols, resulting in the ISI to these samples in the $n$th symbol. 

\begin{figure}
	\centering	
	\subfigure[\scriptsize Before the ISI cancellation]
	{\label{fig:cdf_wISI}\includegraphics[width=0.45\textwidth]{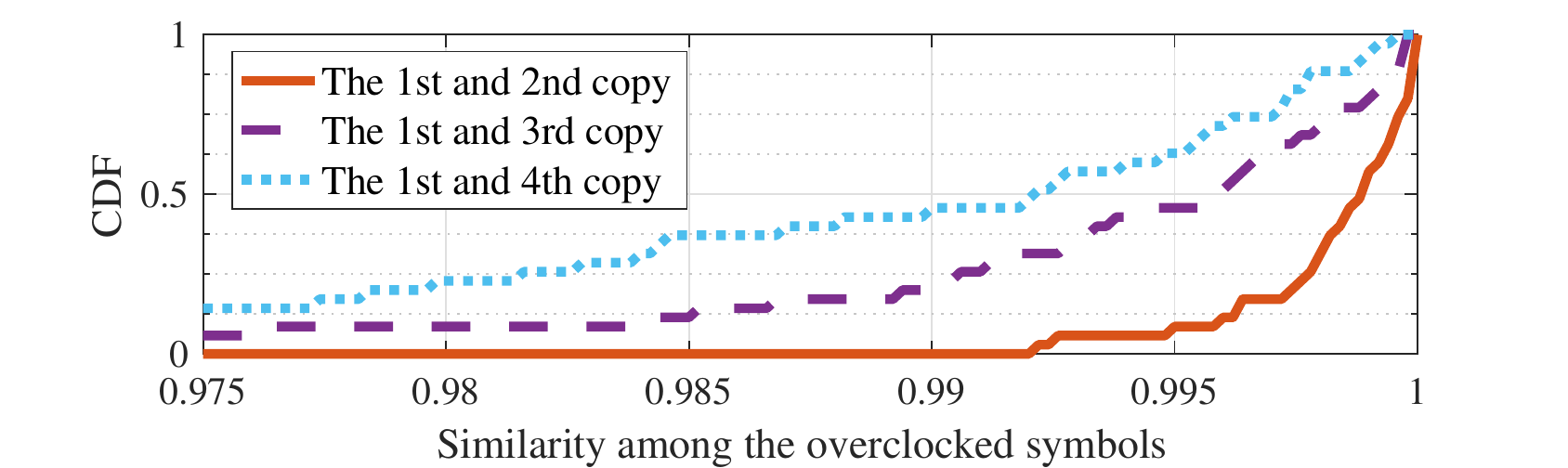}}
	\hspace{-0.2cm}
	\subfigure[\scriptsize After the ISI cancellation]
	{\label{fig:cdf_woISI}\includegraphics[width=0.45\textwidth]{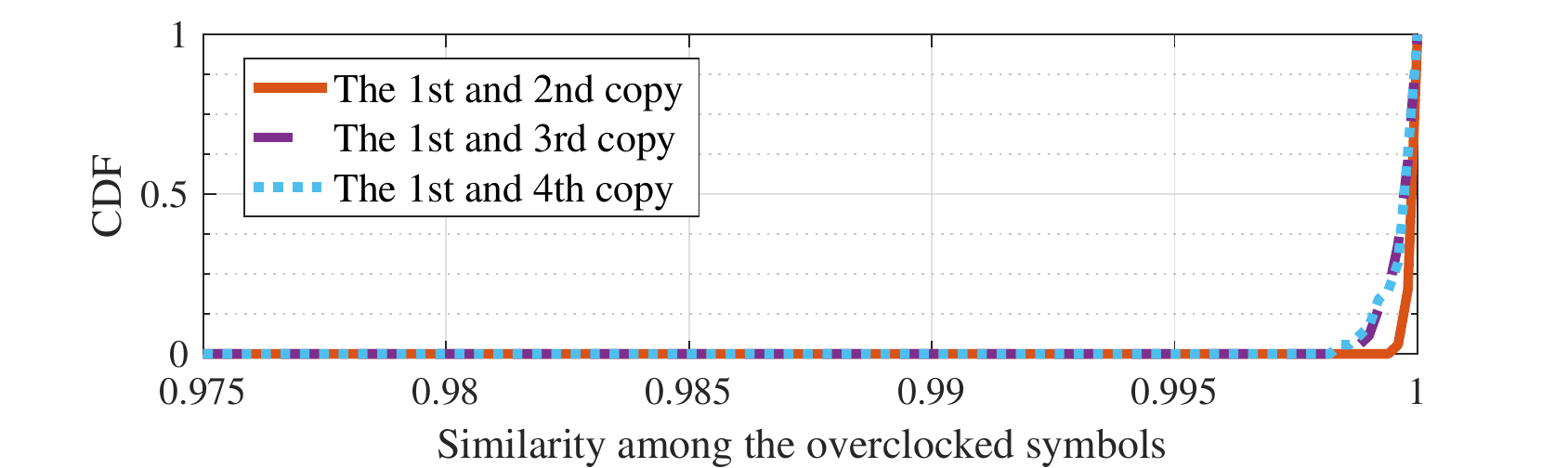}}
	\vspace{0.3cm}
	\caption{CDF for 4-fold overclocked samples in the frequency domain.}
	\label{fig:cdf}
\end{figure}

We investigate how the overclocked signal is impacted by the ISI from the next OFDM symbol.
In particular, we transmit QPSK-modulated packets with random payloads at a high power level, and thus the SNR is higher enough to ignore the effect caused by Gaussian noise. 
We receive all the frames at 4x clock rate and calculate the Pearson correlation coefficient after the FFT and the phase-shift compensation of the time delay. 
Fig.~\ref{fig:cdf_wISI} shows the distribution of similarity, and we observe that the higher this time delay is, the more severe the effect of the ISI is. The difference between the first copy and the interpolated copies decreases the performance of the receiver because the ISI increases the BER of the interpolated copies. After performing the ISI cancellation, overclocked signals will be very similar, as shown in Fig.~\ref{fig:cdf_woISI}. 

\begin{figure}[t]
	\center
	\includegraphics[width=3.5in]{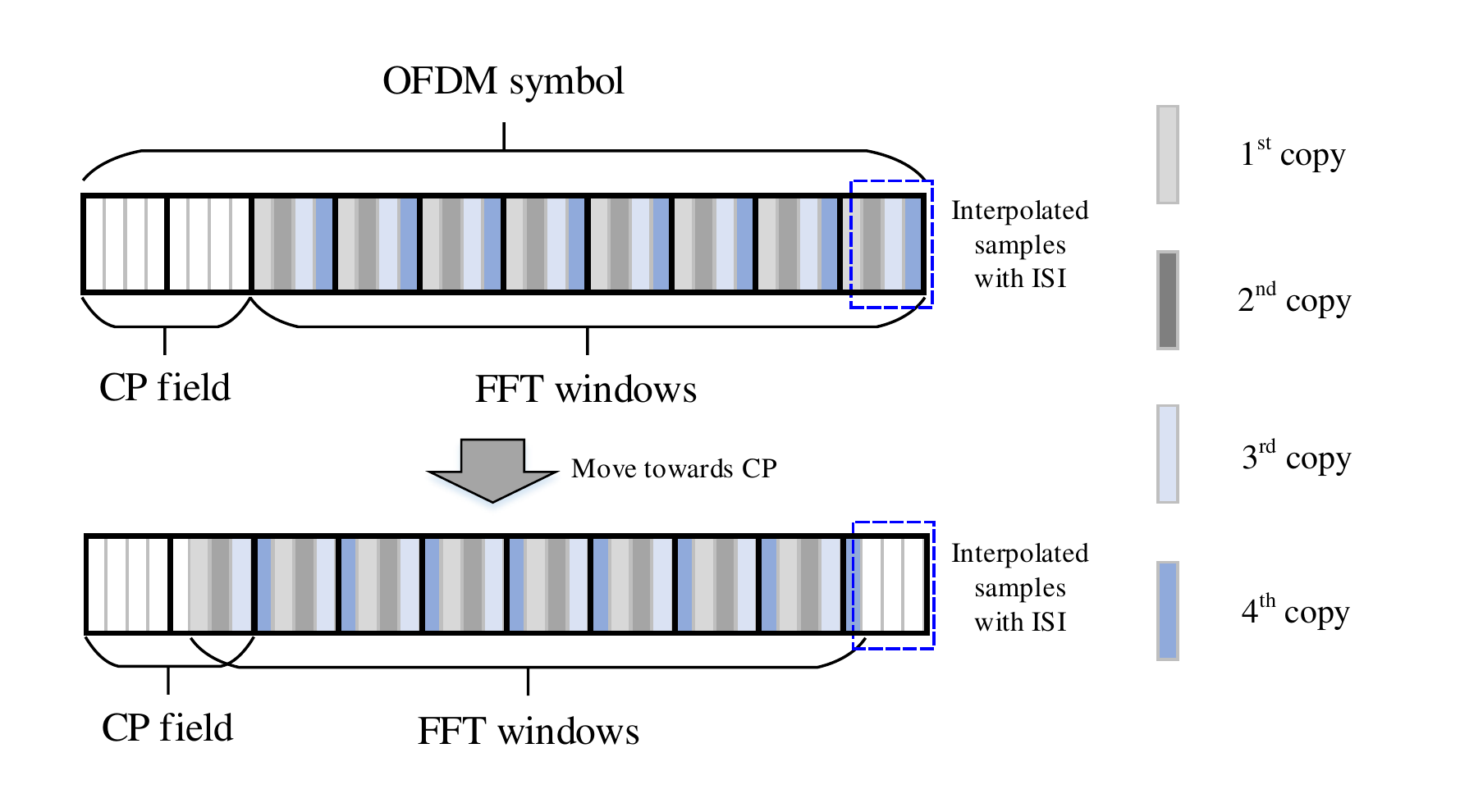}
	\vspace{0.3cm}
	\caption{The shifted FFT windows for \texttt{T-Fi} decoding.}
	\label{fig:move towards cp}
\end{figure}

To reduce the ISI caused by the next OFDM symbol, we avoid using the last few samples in an OFDM symbol for demodulation. Without loss of generality, we illustrate our method by taking $4\times$ clock rate at the receiver as an example. As shown in Fig.~\ref{fig:move towards cp}, there are three redundant copies of the transmitted signal in each OFDM symbol at the receiver. The next OFDM symbol will interfere with the last three interpolated samples of this symbol. Therefore, we simply move the FFT windows towards cycle prefix (CP) to avoid decoding the interfered samples. After the shifting operation, the previous OFDM symbol will not interfere with this OFDM symbol because of the existence of the CP. 
Moreover, considering the multipath effect in a wireless system, we only move $G-1$ samples when we employ $G$-fold clock rate at the receiver to eliminate the ISI. We shift the FFT windows for all copies of received signal both on the LTF and payload. 
Therefore, this operation will neither affect the standard decoding process nor change the phase-shift relation among $G$ copies of the received signal.

After ISI is removed by the shifted FFT window, channel response needs to be understood before packet decoding. Conventional Wi-Fi receivers utilize the two training symbols in LTF to estimate the channel response at each subcarrier. As the data bits encoded on each subcarrier in LTF are pre-known by the receiver, the channel response is obtained by comparing the received data values and the transmitted ones. Based on the estimated channel response, the receiver removes the effect of wireless channel on the following OFDM symbols to extract the transmitted data values.  We extend the ML algorithm to the overclocking case.

Under the overclocking setting, the receiver obtains multiple copies of each data value with different delays. Although we can perfectly model the relation between these copies, our understanding of their noises and interferences is limited. Overclocking introduces disparity between the analog bandwidth and the ADC rate which causes correlation in noise samples~\cite{tepedelenlioglu2004low}. The correlation in noise samples breaks the assumption in the standard decoder.

To effectively leverage the opportunities lying in the redundant samples provided by overclocking, we need to model the effect of residual noises after compensating the phase shifts in redundant samples. Instead of deterministically estimating the channel response and assuming Gaussian distribution for noise samples, our idea is to jointly model the channel response and noise as one mapping function. As such, we retain the information obtained in LTF as much as possible while making fewest assumptions. To this end, we employ kernel density estimation (KDE)~\cite{terrell1992variable}. KDE is a non-parametric data smoothing method to estimate the probability density function (PDF) of a random variable based on a finite number of data samples. We feed the oversampled LTF data values into KDE to generate the PDF of the mapping from the transmitted data value to the received one. This mapping contains the channel response and the noise. As such, we estimate not only the channel response but also the noise distribution at each subcarrier. Since noise samples under normal clock rate follow Gaussian distribution, we use the Gaussian distribution as the kernel in KDE. Specifically, the PDF of the mapping is estimated as
\begin{align}
	& \hat{f}_l(a,\phi) = \notag \\
	& {1 \over GN} \sum_{g=0}^{G-1} \left[{1 \over \sqrt{2\pi}} \exp \left\{-\left({a- \mathcal{A}(Y_g[l]e^{{j 2 \pi gl \over 64G}}-X[l]) \over \sqrt{2}h_a}\right)^2\right\}\right.\notag \\
	& \left.\times {1 \over \sqrt{2\pi}} \exp\left\{-\left({\phi- \Phi(Y_g[l]e^{{j 2 \pi gl \over 64G}}-X[l]) \over \sqrt{2}h_{\phi}}\right)^2\right\} \right],
\end{align}
where $h_{a}(), h_{\phi}$ denote the smoothing parameters, and $\mathcal{A}(\cdot), \Phi(\cdot)$ denotes amplitude and phase of a variable. Note that $Y_g[l]e^{{j 2 \pi gl \over 64G}}$ is the received data value with compensated phase shift. We use all the two OFDM symbols in LTF to obtain a pdf $\hat{f}_l(a,\phi)$ for each subcarrier.

After building the mapping functions based on the LTF symbols, the next step is to combine redundant samples in the payload for decoding. An intuitive method is to compute the average of all redundant samples, and then adopts conventional ML decoder to identify the lattice point in the constellation map with minimal deviation from the average point. The major issue in this method is that the arithmetic average is largely affected by outliers. To address this issue, we jointly consider all the redundant samples in the ML decoder, in which the probability of outliers are used to amortize their impact. In particular, we computing the joint probability of the received redundant samples and identify the lattice point $X[l] \in \{p_1,...,p_K\}$ with the highest likelihood as the transmitted data value. The joint ML decoder identifies the lattice point on subcarrier $l$ by
\begin{equation}
	\hat{X}[l] = \underset{X[l] \in \{p_1,...,p_K\}}{\arg \max} \prod_{g=0}^{G-1} \Pr[Y_g(l)|X(l)].
\end{equation}
The likelihood probability $\Pr[Y_g(l)|X(l)]$ is drawn from the PDF $\hat{f}_l(a,\phi)$ derived by KDE.

\subsection{Protocol Integration and Transceiver Cost} %0.5 pp

\subsubsection{Protocol Integration}
\texttt{T-Fi} redesigns the reception pipeline while completely conforming to Wi-Fi standard for channel access, packet transmission/reception flow. The only departure is that IoT nodes are allowed to use lower power for transmission. Corresponding changes in protocol settings are made to accommodate the low power transmission.

\textbf{Carrier sense.} \texttt{T-Fi} is fully compatible with IEEE 802.11 protocols and can be integrated into any protocol version for channel access and packet transmission/reception. Since IoT devices are allowed to use lower TX power, carrier sense settings need to be amended to sense weak signals. We lower the threshold of Clear Channel Assessment (CCA) according to the amount of change in the TX power. For example, overclocking by 8 folds grants 6.5~dB TX power reduction, and thereby the CCA threshold is tuned to be 6.5~dB lower than the current setting.

\textbf{Rate Adaptation.} Commercial network interface controllers (NICs) employs rate adaptation schemes to adjust MCS modes of each packet based on channel conditions. \texttt{T-Fi} can decode packets of certain MCS modes at lower SNR conditions. Thus, SNR-based rate adaptation schemes can lower the SNR thresholds of each MCS mode to take full advantage of \texttt{T-Fi}. Commercially-adopted rate adaptation schemes such as Auto Rate Feedback (ARF)~\cite{arf} and MadWiFi Ministrel~\cite{ministrel} select MCS mode based on the ratio of ACK frames to previous transmission attempts. In this case, rate adaptation schemes work perfectly in \texttt{T-Fi} and thus requiring no modifications.

\textbf{Coexistence with high-speed Wi-Fi.} \texttt{T-Fi} leverages the power of AP that is used to support the conventional high-speed Wi-Fi such as IEEE 802.11n/ac/ax. It is envisioned by leading companies including Intel, Qualcomm, Huawei, that one AP is capable to simultaneously serve low-power IoT devices and high-speed Wi-Fi nodes. Since IEEE 802.11ah operates at 900~MHz, which has no interference with conventional 2.4/5~GHz Wi-Fi bands. The industry is considering to extend IEEE 802.11ax to support IoT devices at 2.4/5~GHz bands. In this case, the low TX power induced by \texttt{T-Fi} needs to be considered to ensure fairness and avoid interference between IoT devices and conventional high-end devices.

\begin{figure}[t]
	\center
	\includegraphics[width=3.5in]{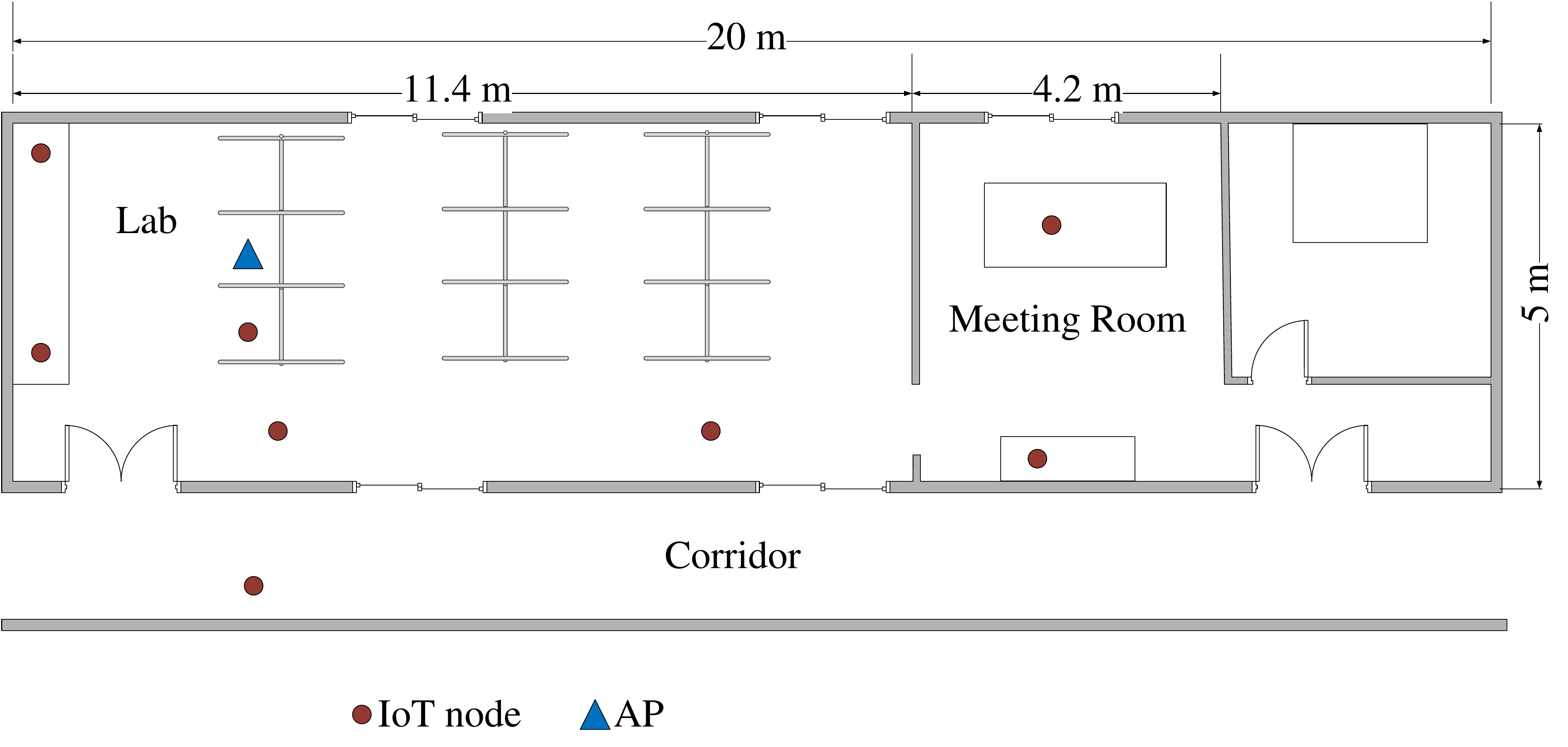}\vspace{0.3cm}
	\caption{Experimental floorplan.} \label{fig:floorplan} 
\end{figure}

\subsubsection{Additional Costs at AP}
The design rationale of \texttt{T-Fi} is to sacrifice the power and computational resources of the AP to support the low-power transmission of IoT devices. To make \texttt{T-Fi} a feasible implementation for today's hardware, the energy cost and decoding complexity are up-bounded by those of IEEE 802.11n/ac/ax.

\textbf{Energy cost.} Basically, an AP stays in one of the four states: packet transmission, packet reception, idle listening, and sleep mode. It has been reported that idle listening consumes 90\% energy for most nodes~\cite{mobicom11emili}. \texttt{T-Fi} uses normal clock rate for idle listening and packet detection, and only switches to overclocking mode for synchronization and packet decoding. Thus, the energy cost induced by overclocking only affects a small portion of the total energy. The digital energy consumption grows linearly with the clock rate, while the analog energy consumption remains the same. Hence, when overclocking by a factor of 8, the total energy cost is still less than twice of the energy cost with standard clock rate. Note that even in the overclocking setting, the clock rate is still less than IEEE 802.11ac/ax (10-80$\times$ clock rates), and thus the energy cost of \texttt{T-Fi} is less than that of IEEE 802.11ac/ax.

\textbf{Extra computational overhead.} The differences between \texttt{T-Fi}'s AP and the normal AP lie in the modification of the conventional modules and an additional phase-shift compensation module, which induce extra computational overhead. We reuse the conventional blocks, such as the timing synchronization block, and thus, the algorithms have comparable time complexity as conventional algorithms as long as AP employs the same clock rate. The extra overhead from the phase-shift module is linear with the FFT length, and is lower than that in the FFT module. Since IoT devices usually need 1-2~MHz bandwidth while APs have 160~MHz, \texttt{T-Fi} can make full use of the redundant resource in the AP with negligible computational overhead increment. Therefore, our design introduces negligible impact on the AP's data processing for both the uplink and downlink transmissions.

\vspace{0.3cm}

\section{Implementation}
\label{sec:implementation}%0.5 pp

\texttt{T-Fi} can be realized in the existing OFDM PHY with no hardware changes. We prototype \texttt{T-Fi} on top of the OFDM implementation on the GNURadio/USRP platform. We implement the entire PHY design specified in Section~\ref{sec:design} directly in the USRP Hardware Drive (UHD). We use USRP B210 nodes connecting to PCs with Intel i7 quad-core processor and 8~GB memory for the testbed setup. Nodes in our experiments are configured to operate in the 2.4-2.5~GHz or 900~MHz range. 

The transmitter is configured to continuously send Wi-Fi packets conforming to IEEE 802.11ah PHY format. 
The symbol duration is 40~$\mu s$. We adopt the legacy PHY layer convergence procedure (PLCP) format of IEEE 802.11ah, where the PLCP preamble consists of 2 STF OFDM symbols and 2 LTF OFDM symbols. There are 10 MCSs starting from 1/2 BPSK to 5/6 256QAM. Due to hardware limitations, we only implement BPSK, QPSK, 16QAM, and 64QAM modulations.
It operates on a 2~MHz or 1~MHz channel, of which 52 subcarriers are configured to carry data values while four subcarriers are pilot tones. This setting conforms to IEEE 802.11ah protocol.

The receiver's pipeline is implemented according to our \texttt{T-Fi} design illustrated in Fig.~\ref{fig:architecture}. Since USRP cannot be overclocked while retaining the same narrow bandwidth in the analog, out-of-band interference and noise are mixed with the oversampled data, which hinders the performance gain of overclocking. We take two steps to minimize the impact of out-of-band interference. First, we scan the 2.4~GHz and 900~MHz bandwidth and select clean channels without nearby interferers. Then, we digitally filter out out-of-band signals immediately after ADC. We modify the synchronization, CFO estimation blocks to incorporate the overclocked samples. The conventional channel estimation block is removed. Instead, we employ a shifted FFT window method to build a channel response and noise model, which is then used to provide Probability Distribution Function(PDF) for a joint ML decoder.

\vspace{0.3cm}

\section{Evaluation}\label{sec:evaluation}%2 pp
In this section, we present a detailed experimental evaluation of \texttt{T-Fi}. Our experiments center around two questions: (i) How much decoding performance improvement can \texttt{T-Fi} provide in real wireless environments under proper overclocking settings? (ii) How much TX power can \texttt{T-Fi} save without compromising the decoding performance? To answer these questions, we conduct a set of experiments to evaluate the performance of synchronization, decoding, the overall packet reception, as well as TX power.

Our goal in this evaluation is to demonstrate that \texttt{T-Fi} can significantly improve the decoding performance at low SNRs and reduce TX power.

\subsection{Experimental Setup}

\begin{figure}[t]
	\center
	\includegraphics[width=3.5in]{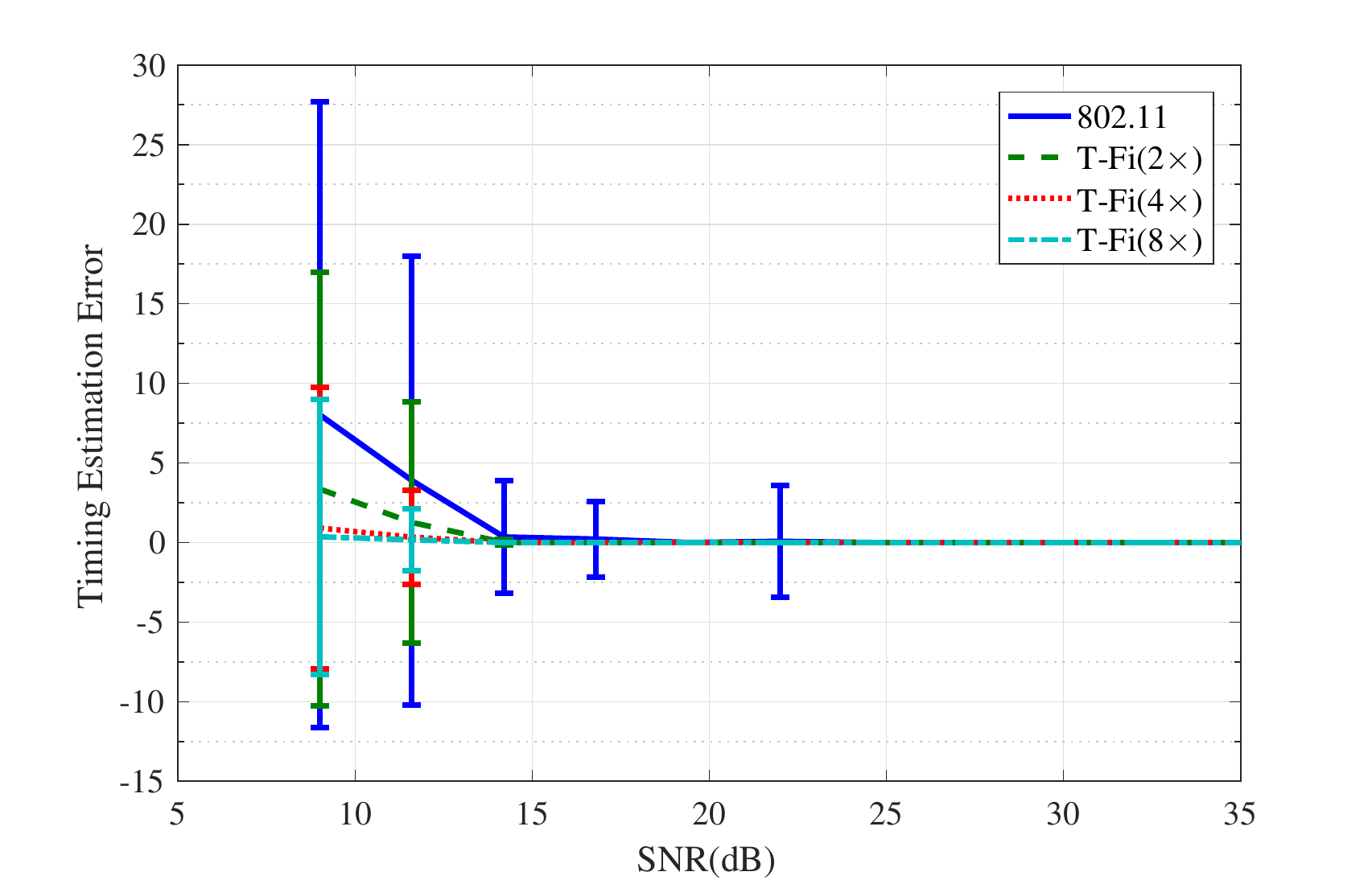}\vspace{0.3cm}
	\caption{Synchronization error.} \label{fig:sync}
\end{figure}

\begin{figure*}
	\centering
	\begin{minipage}[b]{0.49\textwidth}\centering
	\center
	\includegraphics[width=1\textwidth]{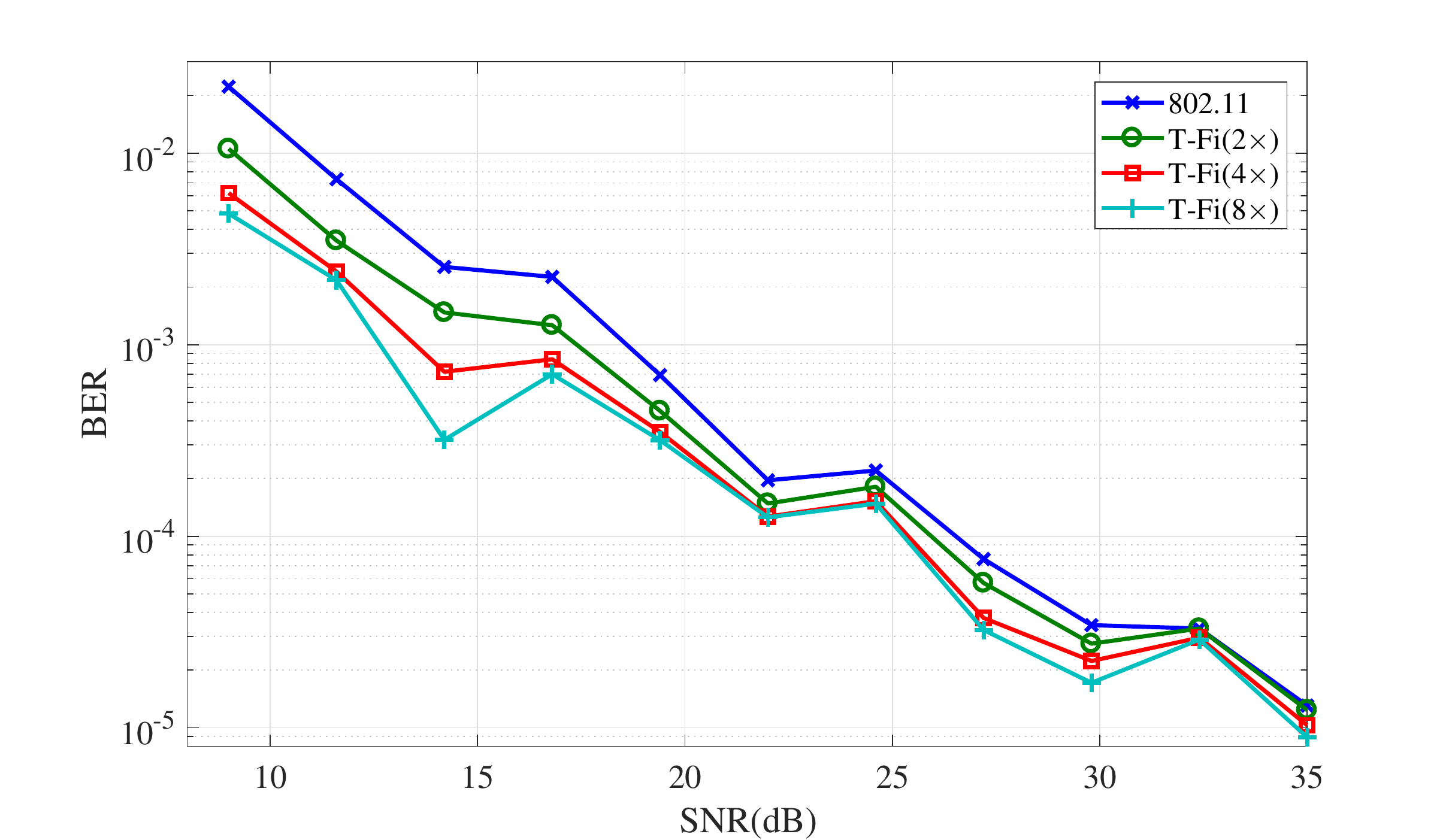}
	\caption{BER under various sample rates.} \label{fig:samplerate}\vspace{0.1cm}
	\end{minipage}\vspace{0.1cm}
	\hspace{-0.1cm}
	\begin{minipage}[b]{0.49\textwidth}\centering
	\center
	\includegraphics[width=1\textwidth]{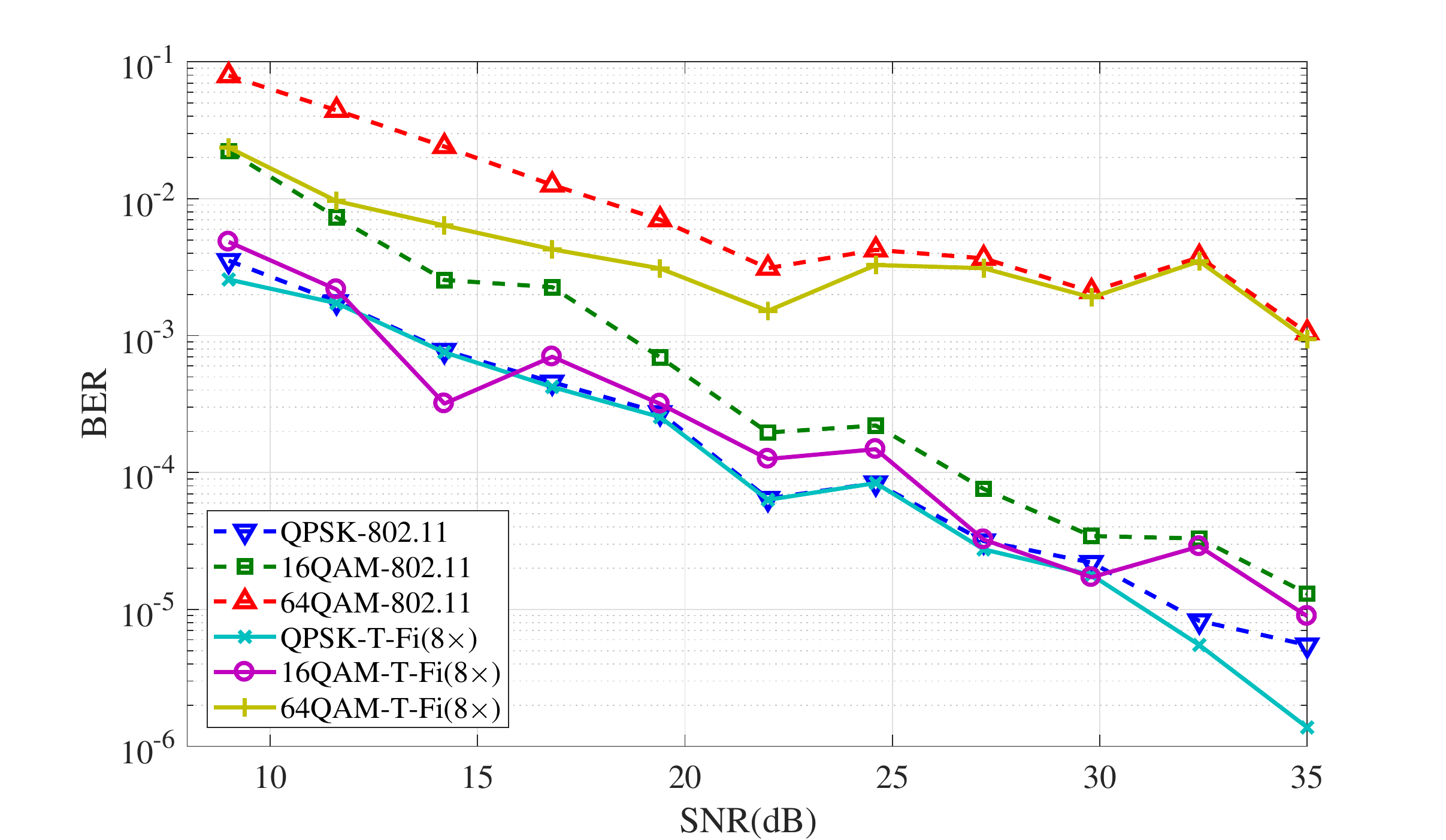}
	\caption{BER for different modulations.} \label{fig:modulation}\vspace{0.1cm}
	\end{minipage}\vspace{0.1cm}
\end{figure*}

We conduct our experiments using USRP B210 nodes deployed in an indoor environment with metal/wood shelves and stone walls. Fig.~\ref{fig:floorplan} illustrates the floorplan of the environment, where nodes are placed in a lab, a meeting room, a hallway, and a corridor. We control the TX power by adjusting the programmable-gain amplifier (PGA) in USRP. As such, we sweep the SNR range from 9~dB to 35~dB. For each setting, a USRP node sends approximate 3000 packets. For each setup, we vary the clock rates of the receiver from the normal clock rate ($1\times$) to overclocked rates ($2\times$, $4\times$, $8\times$). Note that the clock rate of IEEE 802.11ac/ax receiver is $10$-$80\times$. We use the standard IEEE 802.11 packet reception at the normal clock rate as the baseline for comparison.

\subsection{Synchronization}

First, we show that \texttt{T-Fi} addresses the synchronization issue under low SNR conditions. Recall that under low SNR conditions, conventional cross correlation algorithm cannot cancel out noise and yields multiple false peaks. We instead exploit the inherent correlation in the oversampled LFT signals to enhance the correlation property. Since we cannot directly obtain the ground truth of synchronization, we capture raw USRP samples and conduct offline analysis. In particular, we log raw samples received by a USRP node and feed them into an emulator to rehearse the synchronization process. We run the synchronization \texttt{T-Fi} at multiple clock rates (2$\times$, 4$\times$, and 8$\times$ clocking rates) and the standard synchronization algorithm at the normal clock rate (1$\times$ clock rate). Due to the hardware difference between the USRP platform and the commercial Wi-Fi cards, the synchronization results in our experiments might be worse than the results using commercial Wi-Fi cards.

Fig.~\ref{fig:sync} compares the timing estimation error, which is measured in number of samples. Error bars show the standard deviation of time synchronization offset. Under all SNR conditions, synchronization error diminishes with the increment of clock rates. The standard synchronization algorithm works well for SNR$>15$, while the synchronization error grows substantially when SNR is low (9~dB - 12~dB). Empowered by overclocking, \texttt{T-Fi} largely reduces the synchronization error. When SNR=9~dB, the average synchronization error at $8\times$ clock rate is 0.91, which is merely 13\% of the average error at the standard clock rate. These results confirm the merit of using over-clocking to do time synchronization. 

\subsection{Decoding}

\begin{figure*}
	\centering
	\begin{minipage}[b]{0.66\textwidth}\centering
		\subfigure[\scriptsize Different sample rates]
		{\label{fig:loc_rate}\includegraphics[width=0.5\textwidth]{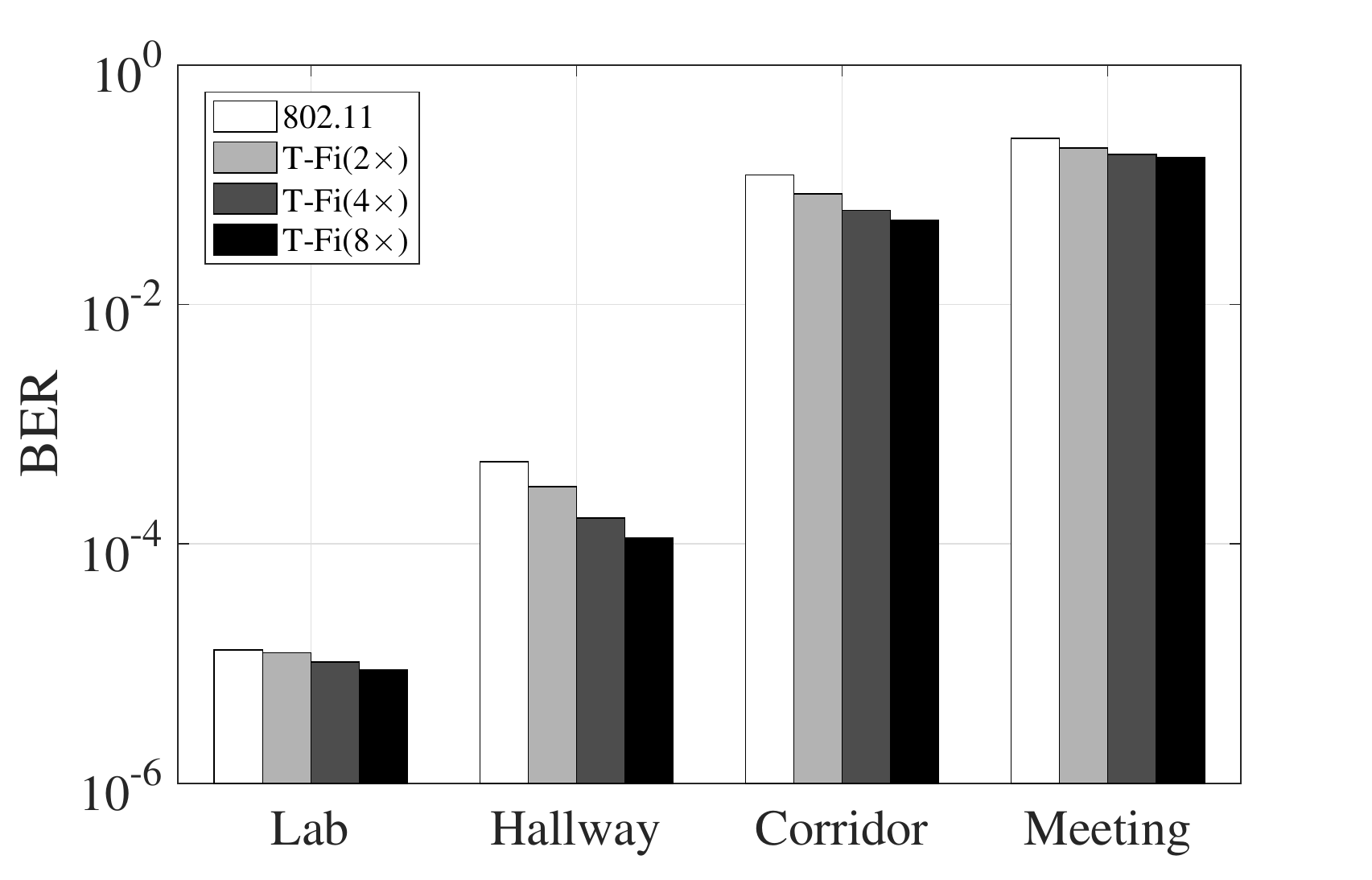}}\hspace{-0.2cm}
		\subfigure[\scriptsize Different modulations]
		{\label{fig:loc_mod}\includegraphics[width=0.5\textwidth]{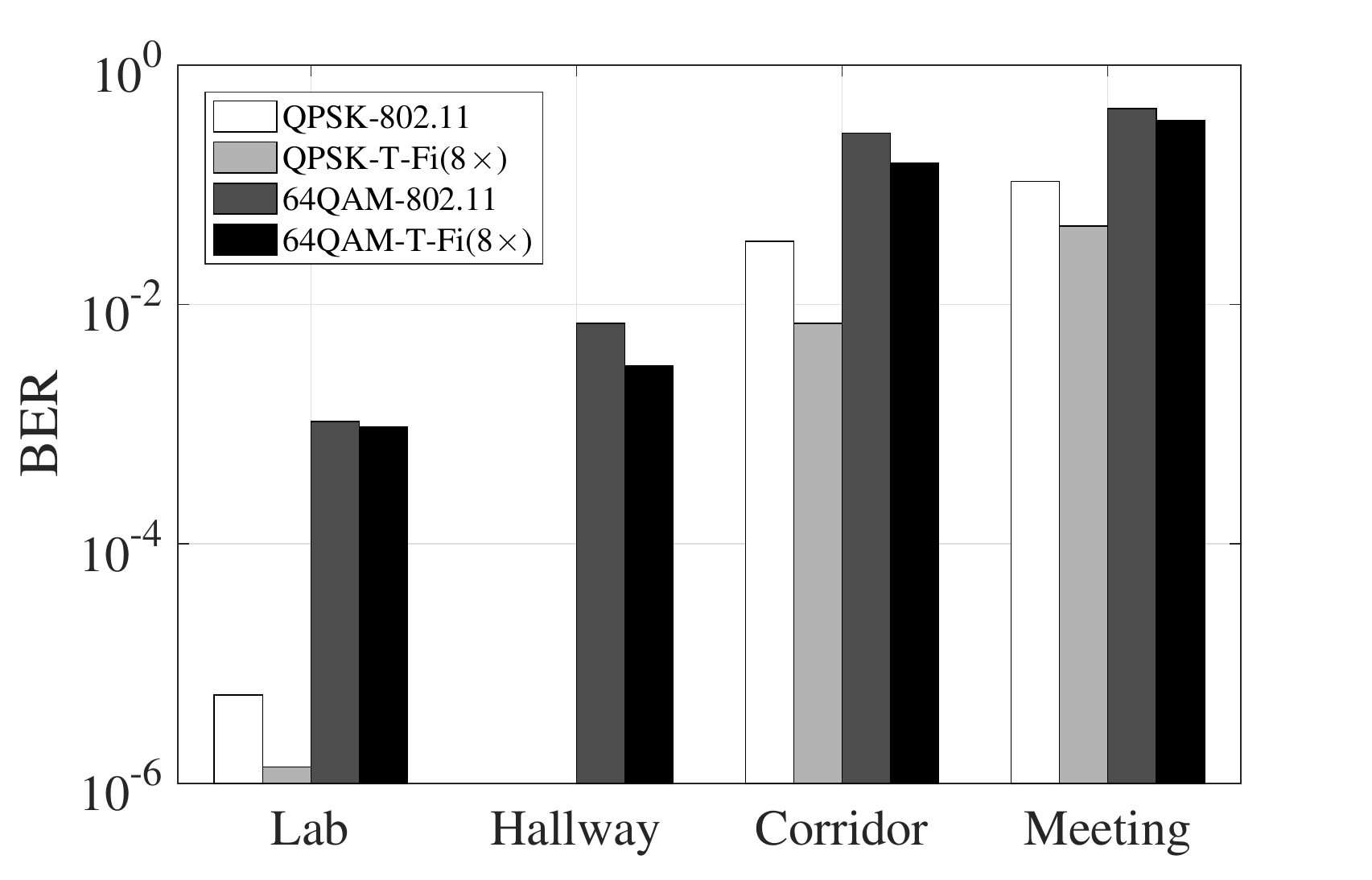}}\vspace{0.3cm}
		\caption{BER in different locations.}
		\label{fig:loc}
	\end{minipage} 
	\vspace{0.2cm}
	\hspace{-0.2cm}
	\begin{minipage}[b]{0.33\textwidth}\centering
		\includegraphics[width=1\textwidth]{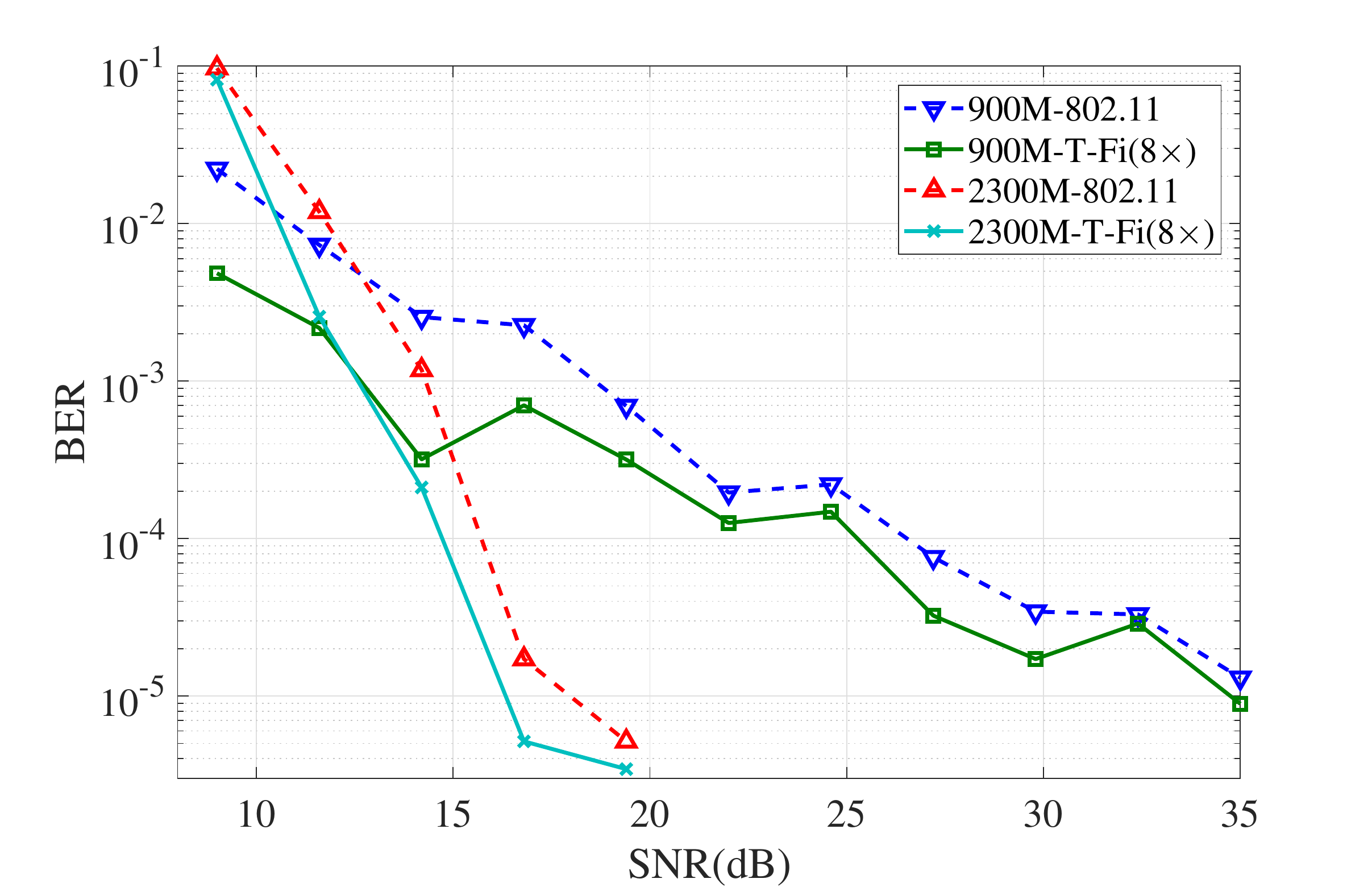}
		\vspace{0.2cm}
		\caption{BER under different carrier frequencies.} \label{fig:freq}\vspace{0.01cm}
	\end{minipage} 
	\vspace{0.2cm}
\end{figure*}

%2. decode: (use the same sync algorithm)
\textbf{Decoding performance under various SNR conditions.} Our second set of experiments evaluates the decoding performance at different MCS modes under various environments and SNR conditions. To focus on the decoding performance, we use the same synchronization algorithm at the same clock rate for all decoders. In particular, we set $8\times$ clock rate to receive packets, and perform synchronization and CFO compensation. Unless otherwise stated, USRP nodes operate at 900~MHz, which is the frequency band specified in IEEE 802.11ah.

Then, we digitally downscale the clock rate to the corresponding rates (1$\times$, 2$\times$, 4$\times$, 8$\times$) for packet decoding. Fig.~\ref{fig:samplerate} shows the bit error rate (BER) of \texttt{T-Fi} and receiver under various SNR conditions. We vary the PGA of the USRP sender as well as the transmission distance to induce different SNR. Packets are modulated using 16QAM. Under high SNR conditions (SNR$>$15~dB), \texttt{T-Fi} achieves marginal gain over the standard 802.11 receiver. Below the 15~dB threshold, BER of the standard 802.11 receiver increases substantially, as the SNR range is lower than the minimal threshold required for decoding 16QAM-modulated packets. Under low SNR conditions, \texttt{T-Fi} achieves BER$<1.1\%$ at $2\times$ clock rate and BER$<0.5\%$ at $8\times$ clock rate. The BER gain comes from the redundancy between oversampled data, which is exploited in \texttt{T-Fi} to amortize the effect of random noise samples.

The average degradation of BER at all SNRs is 28.95\%, 26.36\%, and 15.41\% when moving clock rate from $1\times$ to $2\times$, $2\times$ to $4\times$, and $4\times$ to $8\times$, respectively. The diminishing returns when the clock rate increases are caused by the imperfect hardware and environmental interference. Specifically, the roll-off of the imperfect lowpass filter~[20] leads to performance degradation at the channel border. Likewise, the nonlinearity of the lowpass filter leads to the peak-to-average power ratio (PAPR) issue at high transmission power. The hardware defect causes signal distortion and increases the BER. Since all copies in \texttt{T-Fi} pass through the same hardware, it is difficult to eliminate its impact. On the other hand, the environmental interference causes the noise is not completely independent and produces bursty errors~[21] during the transmission, resulting in diminishing returns at high clock rate.

We further analyze the average BER below 14.2~dB and observe that the BER decreases 48.99\%, 41.12\%, 29.19\% when doubling the clock rate for $1\times$, $2\times$, and $4\times$ clock rate, respectively. The average performance gain below 14.2~dB is higher than that at all SNRs. The reason is that the BER in the experiment is related to both the synchronization and decoding module in the reception pipeline, while the timing synchronization module works well at the high clock rate. Therefore, further increase in clock rate at high SNR does not bring gain from these modules to the whole system.

In Fig.~\ref{fig:modulation}, we compare the decoding performance under different modulation schemes. The clock rate of \texttt{T-Fi} receiver is $8\times$ of the standard 802.11 receiver. As expected, for all modulations demonstrated, \texttt{T-Fi} outperforms the standard 802.11 receiver. We also observe that \texttt{T-Fi} substantially reduces BER compared to the standard receiver for high-order modulations. When SNR=9~dB, the BER of \texttt{T-Fi} is merely 25\% of the standard receiver's BER. The reason is that high-order modulations are more error-prone under low SNR conditions, which leaves more room for overclocking to correct these errors.

Another observation is that the decoding performance of \texttt{T-Fi} for 64QAM (16QAM) is comparable to that of the standard receiver for 16QAM (QPSK). This result implies that \texttt{T-Fi} can use higher-order modulations for packet transmission compared to standard Wi-Fi, and thus delivers roughly $4\times$ data rate. Hence, we can envision a new dimension of the \texttt{T-Fi} is to improve the throughput in IoT Wi-Fi.

\textbf{Impact of wireless environment.} As analyzed in Section~\ref{sec:overclock}, the decoding performance depends on channel response and noise. In our previous experiments, all nodes are in the same lab, and we focus on the decoding performance under different SNR conditions. In this experiment, we evaluate the impact of wireless environments, which lead to different multi-path fading and shadowing. We repeat the previous set of experiments at three additional locations in the same building but with significantly different propagation environments. The locations of IoT nodes and the AP include a hallway in the lab, a corridor outside the lab, and a meeting room next to the lab, as illustrated in Fig.~\ref{fig:floorplan}. We use a fixed PGA to set the same TX power for all locations.

Fig.~\ref{fig:loc_rate} compares the decoding performance under different clock rates at different locations. For all locations, \texttt{T-Fi} outperforms the standard 802.11 decoding, and higher clock rate \texttt{T-Fi} receivers achieve lower BER than the lower clock rate ones. Compared to the standard receiver, \texttt{T-Fi} at $8\times$ clock rate reduces BER to 56\%-23\% across all locations. \texttt{T-Fi} yields higher performance gain in the hallway and the corridor. The reason behind this is that the BER in these two locations falls in a sweet range where there is room for BER improvement while noise can be amortized by exploiting overclocking. The results confirm that \texttt{T-Fi} achieves stable performance gain over a wide range of wireless environment. Fig.~\ref{fig:loc_mod} further shows the BER performance for different modulation schemes at the four locations. Though the BER of different modulations varies significantly across different locations, \texttt{T-Fi} achieves steady BER gain over the standard receiver.

%2.5 BER vs SNR (900MHz, 2.4G)

\begin{figure}[t]
	\center
	\includegraphics[width=3.5in]{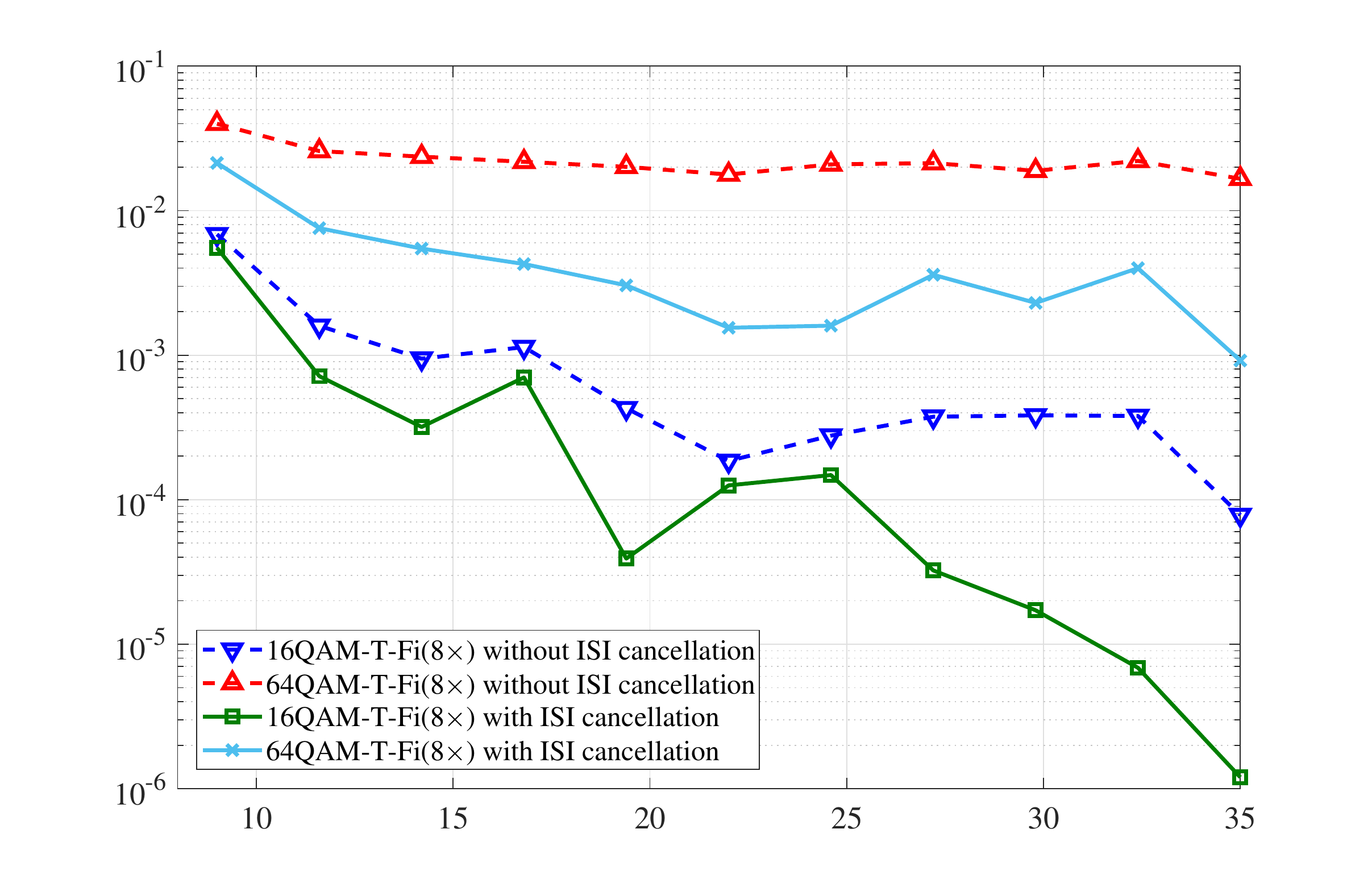}
	\caption{BER with and without the ISI cancellation.}
	\vspace{0.3cm}
	\label{fig:BER_wo_ISI}
\end{figure}

\begin{figure*}
	\centering
	\begin{minipage}[b]{0.49\textwidth}\centering
		\center
		\includegraphics[width=1\textwidth]{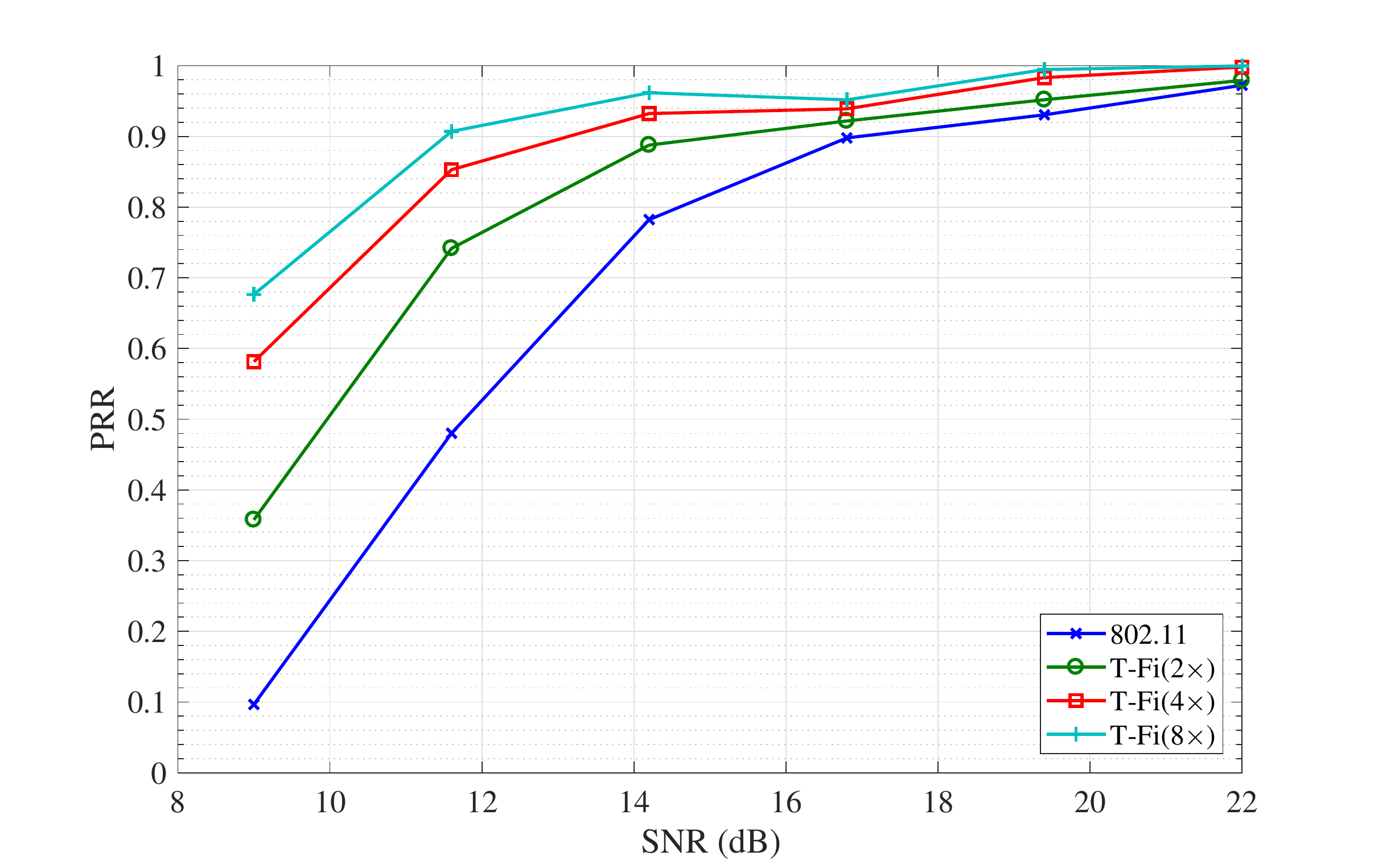}
		\caption{PRR under various SNR conditions.} \label{fig:prr_snr}
	\end{minipage}\vspace{0.1cm}
	\hspace{-0.1cm}
	\begin{minipage}[b]{0.49\textwidth}\centering
		\center
		\includegraphics[width=1\textwidth]{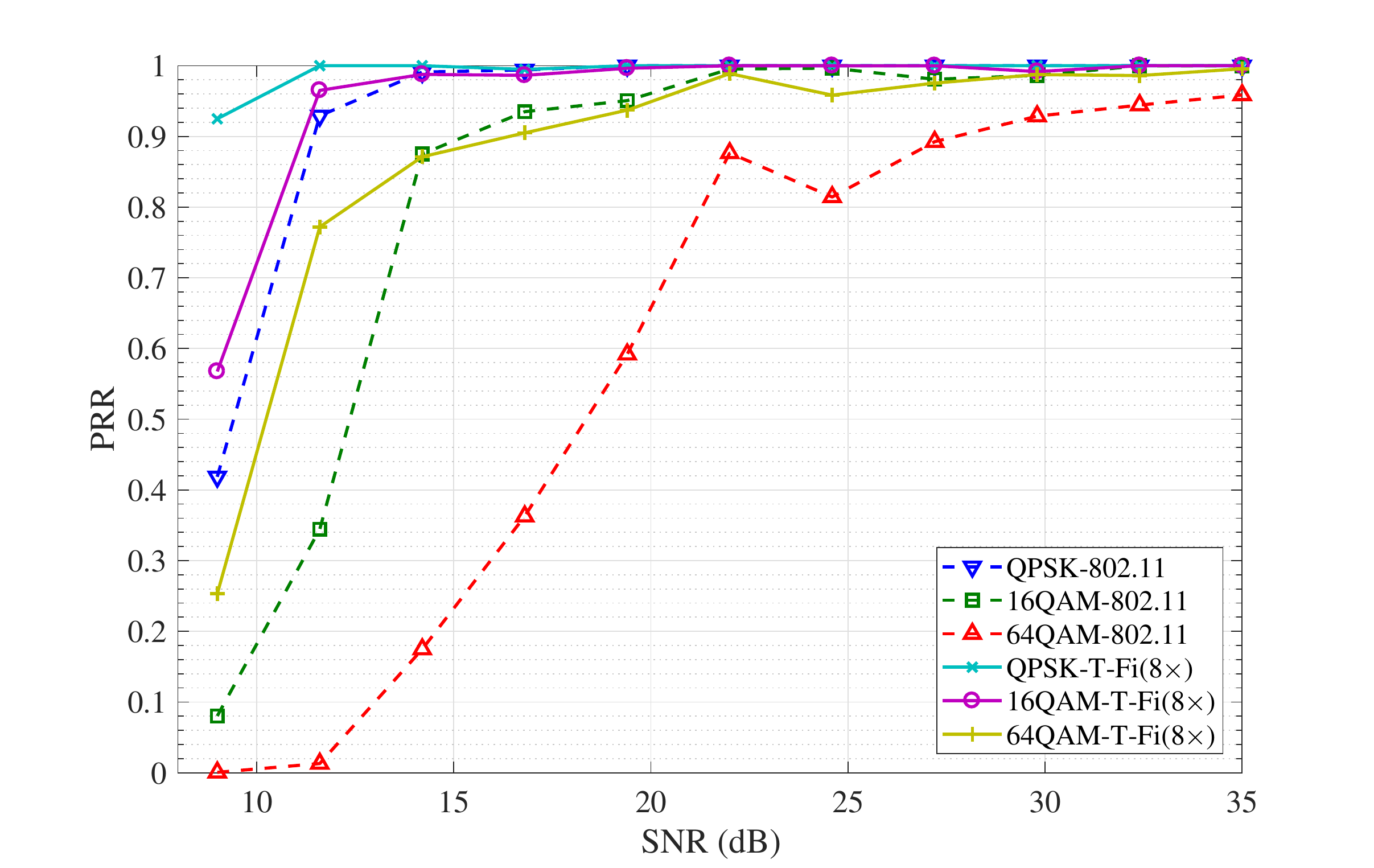}
		\caption{PRR for different modulations.} \label{fig:prr_mod}
	\end{minipage}\vspace{0.1cm}
\end{figure*}

\textbf{Impact of carrier frequency.} Previous experiments use 900~MHz as the carrier frequency, which is the band specified by IEEE 802.11ah. Recently, several leading companies have been exploring the possibility of using 2.4~GHz for IoT transmissions. To meet this potential demands, we now evaluate the BER performance in both carrier frequencies under various SNR conditions.

Fig.~\ref{fig:freq} shows the BER for 16QAM decoding. We have the following observations: (i) The BER performance at both frequencies has the same trend and is comparable in most cases; (ii) \texttt{T-Fi} achieves lower BER compared to the standard receiver at both frequencies. The results validate the feasibility of \texttt{T-Fi} at both carrier frequencies.

\textbf{Impact of ISI.} In the previous evaluations, we focus on the decoding performance without the ISI. As analyzed in Section~\ref{sec:overclock}, the decoding performance depends on both the Gaussian noise and the ISI. To evaluate the effect of ISI, we set various SNR conditions at different MCS modes which are similar to the previous experiments, and we transmit random bits in the payload and compare both the decoding performance before and after the ISI cancellation. The results are shown in Fig.~\ref{fig:BER_wo_ISI}.

With the shifted FFT windows, the BERs are significantly decreased for both the 16QAM and 64QAM modulation schemes. Considering the BER at 9dB, for example, it can be observed that the decoding performance of QAM modulation is improved. \texttt{T-Fi} at $8\times$ clock rate reduces BER to 79\% and 54\% for 16QAM and 64QAM, respectively. The reason is that the higher modulation scheme suffers more severe ISI. 

\subsection{Overall Packet Reception}
Previous experiments evaluate individual components independently. We now evaluate the overall packet reception performance. We set different clock rates for packet reception, and employ \texttt{T-Fi} or standard 802.11 reception pipelines for synchronization, CFO compensation, and packet decoding. We measure the packet reception ratio (PRR), which is the ratio of the number of correctly decoded packets to the total number of packets transmitted. The packet length is set to be 100 Bytes, and the coding rate is set to be 1/2 for all modulations.
{{To ensure a better performance, we also consider the ISI cancellation algorithm in the experiments. }}
 
Fig.~\ref{fig:prr_snr} shows the PRR of \texttt{T-Fi} and the standard receiver under various SNR conditions. Packets are modulated using 1/2 16QAM. 
{{For SNR=14~dB and above, standard receiver and \texttt{T-Fi} achieves good PRR of at least 77\% and 95\%, respectively, and the performance gap between \texttt{T-Fi} and the standard receiver diminishes with higher SNR. At 12~dB SNR, the standard receiver yields merely 54\% PRR, while \texttt{T-Fi} achieves 75\%-92\% PRR with $2\times$-$8\times$ clock rate. For the lowest SNR (9~dB) measured in the experiment, the standard receiver fails to decode 97\% of all packets, while \texttt{T-Fi} still achieves a noticeable improvement (PRR = 67.6\% for $8\times$ clock rate). }}
The results show that \texttt{T-Fi} can significant improve the packet decoding under poor SNR conditions.

By comparing the PRR between \texttt{T-Fi} at different clock rates, we find that \texttt{T-Fi} at $2\times$ and $4\times$ clock rates yields at most 33\% and 22\% PRR improvements when doubling the clock rate, while \texttt{T-Fi} at $8\times$ improves only 10\% when doubling the clock rate from $4\times$ clock rate. 
It reveals that most redundancy in overclocked samples can be exploited at lower clock rates. It implies that we can get the most of the redundancy gain under affordable overclocking settings. As the clock rate gap between IoT devices and APs is 10 to 80 times, $8\times$ overclocking is a reasonable setting to achieve desirable performance under the constraint of AP's hardware capability.

Fig.~\ref{fig:prr_mod} illustrates the PRR for various modulation modes. The results are very promising for all modulation modes. We observe that the standard receiver suffers a sharp drop at much higher SNR compared to \texttt{T-Fi}. \texttt{T-Fi} even achieves better decoding performance for 64QAM (16QAM) than the performance for 16QAM (QPSK) using the standard receiver, which is consistent with the BER results in Fig.~\ref{fig:modulation}. It demonstrates that \texttt{T-Fi} has the potential to improve the data rate under low SNR conditions.

\subsection{Transmission Power Reduction}

\begin{figure*}[t]
	\subfigure[\scriptsize Different sample rates]
	{\label{fig:power_rate}\includegraphics[width=2.35in]{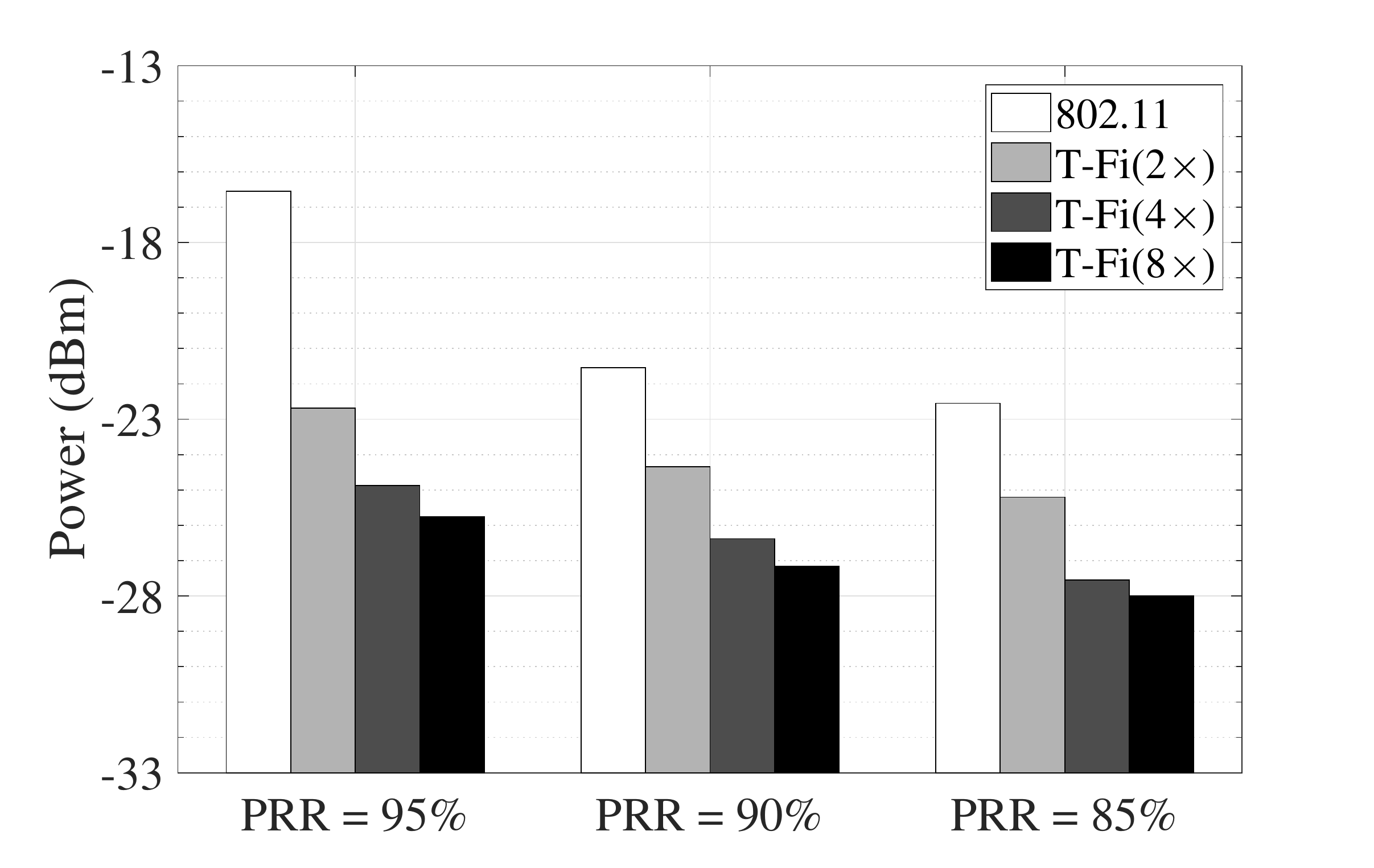}}\hspace{-0.3cm}
	\subfigure[\scriptsize BPSK and QPSK]
	{\label{fig:power_psk}\includegraphics[width=2.35in]{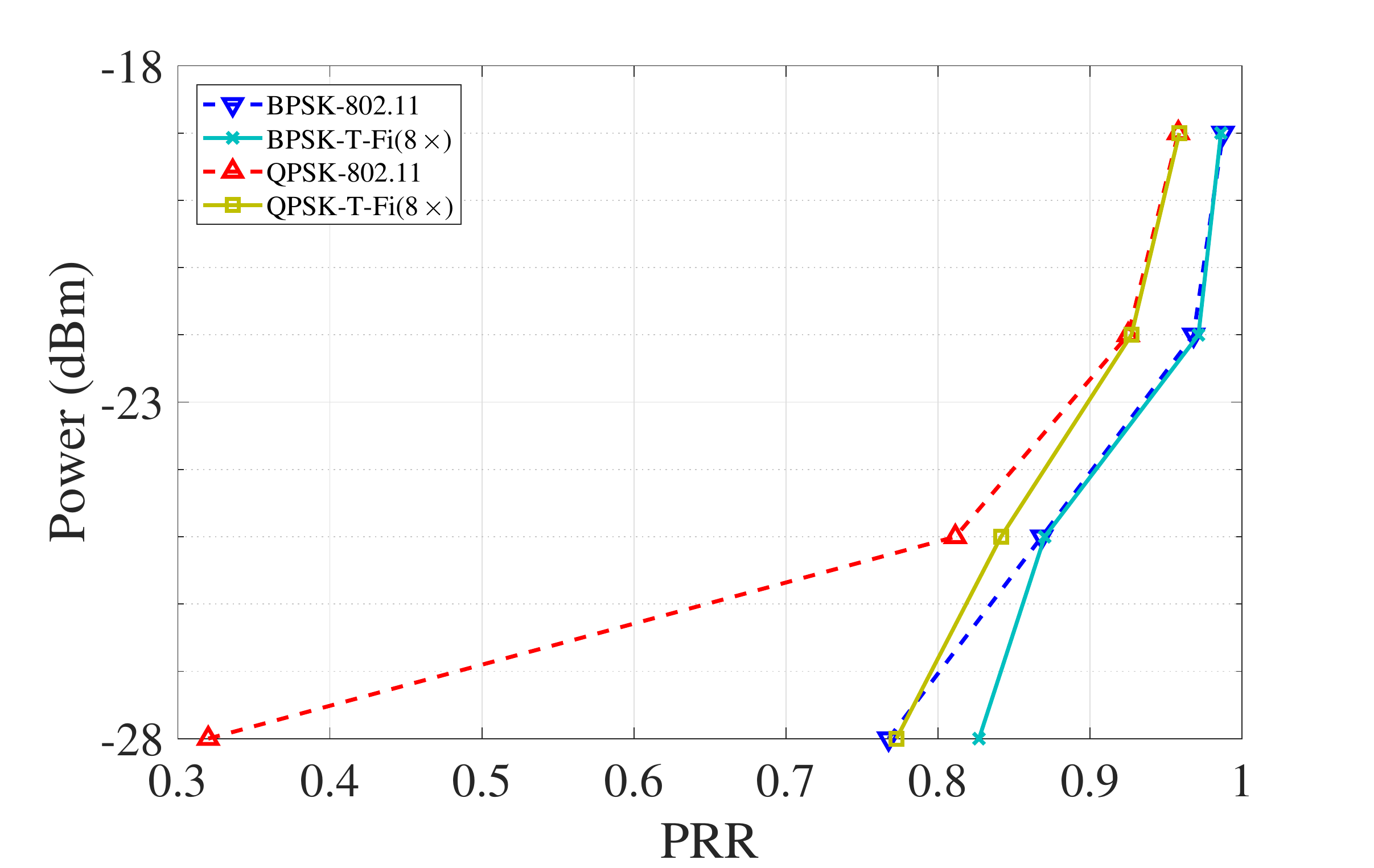}}\hspace{-0.3cm}
	\subfigure[\scriptsize 16QAM and 64QAM]
	{\label{fig:power_qam}\includegraphics[width=2.35in]{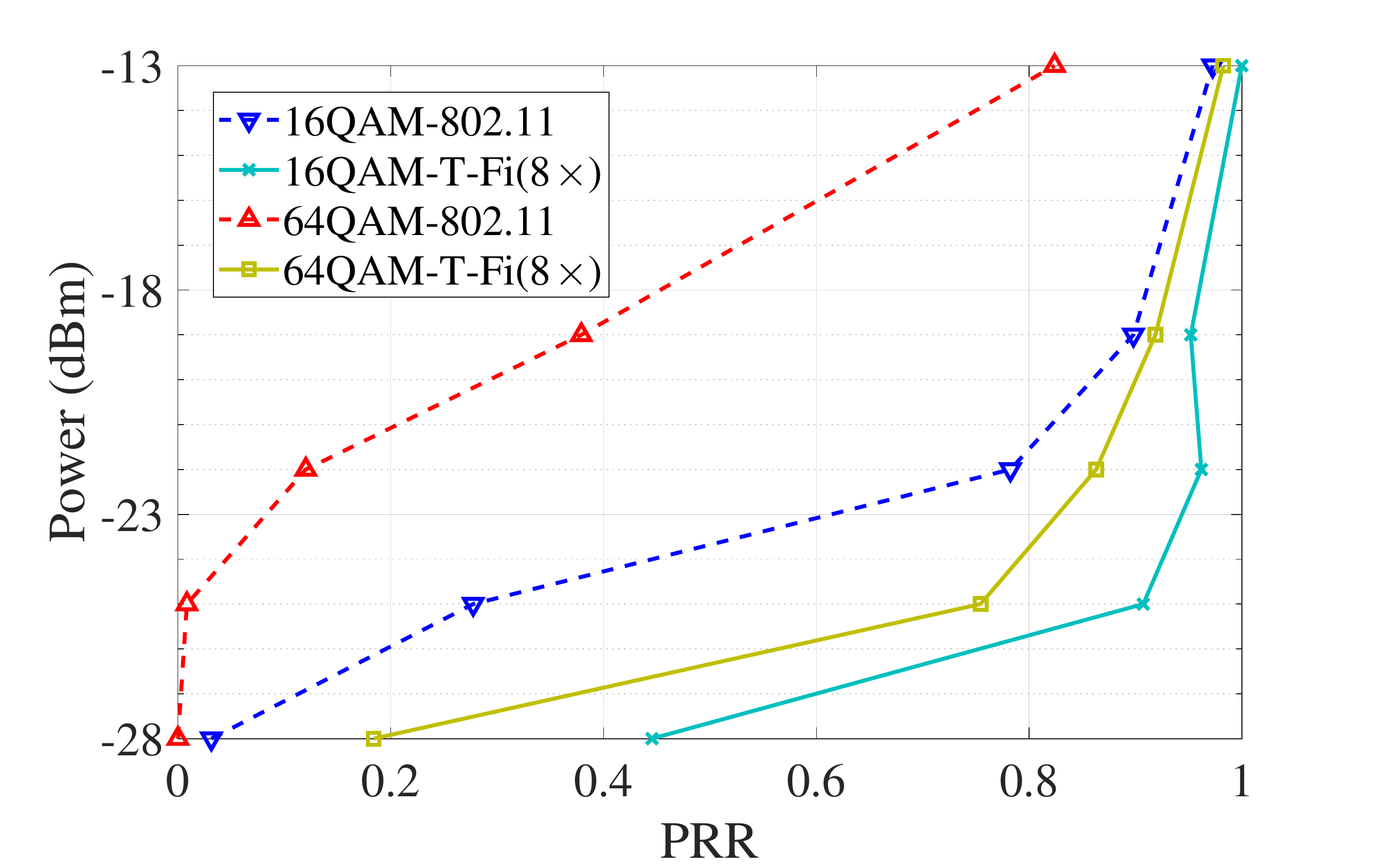}}\hspace{-0.3cm}
	\caption{Transmission power reduction.}\vspace{0.3cm}
	\label{fig:power}
\end{figure*}

Finally, we evaluate the benefits of \texttt{T-Fi} in terms of TX power reduction compared to the standard 802.11 receiver. We use the same setting as in the overall packet reception experiments to evaluate the overall power reduction merits. At each location, we gradually vary the TX power by tuning the PGA in USRP, which lead to different SNRs and thus different PRR. Transmission power corresponding to each value of PGA is identified by checking the datasheet~\cite{b210}. We set target PRR and compare the minimal TX power of different receivers that satisfies the target PRR.
{{The ISI cancellation algorithm is implemented as illustrated in section~\ref{sec:decoding}}.}

Fig.~\ref{fig:power_rate} compares the TX power under various clock rates. \texttt{T-Fi} at $2\times$ ($8\times$) clock rate consumes 43.65\% (22.65\%) of the TX power as used by the standard receiver. \texttt{T-Fi} at $2\times$ and $8\times$ clock rates can save up to 75.62\% and 88\% of the TX power compared to the standard receiver, respectively. The TX power reduction comes from the decoding improvement that allows lower SNR for packet transmission. As the power amplifier dominate the total power consumption of Wi-Fi radios~\cite{qiao2003miser,jung2002energy,baiamonte2006saving,bruno2002optimization,carvalho2004modeling,chen2008edca,ergen2007decomposition,garcia2011energy}, the reduction of TX power can effectively save energy for IoT devices.
For example, the power consumption of MAX2831~\cite{maxim}, which is a common Wi-Fi chipset, reaches 919.3~mW. According to the relation between the total power and the transmission power of the Wi-Fi chipset~\cite{MAX2831_PA}, \texttt{T-Fi} can reduce the energy consumption of this chipset by up to 39.36\%.

We further take a deeper look at the power reduction under different modulation modes in Fig.~\ref{fig:power_psk} and Fig.~\ref{fig:power_qam}. In Fig.~\ref{fig:power_psk}, we observe that there is marginal gain for BPSK. This is because in our lowest power setting BPSK performs well using the standard receiver and thus there is no much room for improvement. In Fig.~\ref{fig:power_qam}, we observe that the slopes of \texttt{T-Fi} are much sharper than those of the standard receiver. It implies that with only slight compromise in PRR \texttt{T-Fi} can yield a large amount of power reduction. For 64QAM, \texttt{T-Fi} reduces TX power from -13~dBm to -25~dBm (93.7\% power reduction) at cost of only 6.52\% PRR drop. We also observe that \texttt{T-Fi} reduces much more TX power for QAM-modulated packets than PSK-modulated packets. The reason behind this observation is that QAM-modulated packets are less robust under poor SNR conditions, and thus overclocking can largely add redundancy to facilitate lower power transmission. 

We finally analysis the power gain of oversampling at the receiver. In theory, the absolute value of the power gain is equivalent to the absolute value of the power reduction at the transmitter. We can get 3~dB power gain when doubling the clock rate. However, the average power gain in the experiments is 3.44~dB, 5.21~dB, and 5.69~dB, respectively. The experimental result is different from the theoretical result for two reasons. First, the performance of the whole system is impacted by other modules, leading to the power gain larger than 3~dB, while the analysis only considers the performance of the decoding module. Second, the noise does not completely obey Gaussian distribution, resulting diminishing returns at the high clock rate.

\vspace{0.3cm}

\section{Related Work}
\label{sec:relatedwork}%0.5 pp

\textbf{IoT Wi-Fi standard.} IEEE 802.11ah, known as Wi-Fi HaLow~\cite{wifihalow}, is the first and only Wi-Fi protocol dedicatedly designed for IoT. The target of this IoT Wi-Fi protocol is to utilize the Wi-Fi technology to provide low power and long range transmission for the emerging low-end IoT devices. It reuses the OFDM PHY as in IEEE 802.11ac, but provides a bundle of low power features, particularly at the MAC layer. Research efforts have focused on enhancing the MAC layer~\cite{park2014enhancement,liu2013power}. Our goal is to design a transceiver architecture for IoT Wi-Fi networks that (i) completely conforms to the IoT Wi-Fi standard without any protocol modifications, and at the same time (ii) take full advantage of AP to further reduce the TX power of IoT devices. Our compliant design can be seamlessly integrated into the existing IoT Wi-Fi protocol. Existing Wi-Fi power models~\cite{qiao2003miser,jung2002energy,baiamonte2006saving,bruno2002optimization,carvalho2004modeling,chen2008edca,ergen2007decomposition,garcia2011energy} show that power amplifier operations dominate the total power consumption of Wi-Fi radios. Thus, the saved power budget by our design can facilitate lower power or longer range transmission, which fully aligns with the targets of IoT Wi-Fi protocols.

\textbf{Energy-efficient protocol.} Energy efficient mechanisms have been studied at all layers of 802.11 stack. Starting from PHY, Qiao et al.~\cite{qiao2007interference} pre-compute optimal TX power for each frame. At MAC layer, backoff~\cite{jung2002energy,baiamonte2006saving} or contention parameters~\cite{chen2008edca,garcia2011energy} are tuned to reduce energy waste. 
A new polling-based MAC protocol is proposed in UPCF~\cite{upcf2006} to save energy for multimedia applications. Serrano et al.~\cite{PerFrame2014} propose an in-depth understanding of the energy consumption in different types of 802.11 devices for both UDP and TCP traffic. 
Many approaches rely on putting Wi-Fi radios into sleep mode to save unnecessary energy consumption while being idle. 

IEEE 802.11 defines a mechanism called Power Saving Mode (PSM)~\cite{psm}, which allows devices to enter a low-power sleep and only wake up every 100~ms to receive AP beacons when they are in the inactive state. Based on the standard PSM, many variants have been proposed to enhance it. Dynamically adjusting sleeping periods have been proposed in ~\cite{mobicom02psm,anand2005self} to better fit the traffic patterns. $\mu$PM~\cite{mobisys08micropsm} extends PSM to frame-level sleeping by powering down light-traffic nodes between individual frame intervals. Capnet~\cite{mobisys10catnap} extends the frame-level sleeping mechanism by jointly considering data availability from wide-area networks.
OPEM~\cite{poem2016} leverages the bandwidth asymmetry between the WWAN and Wi-Fi interfaces to reduce energy consumption of APs.

Another class of energy-efficient protocols adopts a more drastic mechanism: rather than entering a sleeping mode, they completely turn off Wi-Fi cards in a duty-cycle mode. Cell2Notify~\cite{agarwal2007wireless} leverages the always-on cellular radio to turn on Wi-Fi radio whenever the is incoming data. Instead of using cellular radios, Shih et al.~\cite{shih2002wake} design a low-energy control channel to check pending data and turn off radios when there is no traffic. Blue-Fi~\cite{ananthanarayanan2009blue} and FreeBee~\cite{kim2015freebee} leverage low-power Bluetooth radios to discover Wi-Fi networks and only turn on Wi-Fi interfaces after identifying available Wi-Fi networks. DopplerFi~\cite{DopplerFi} extends the discovery algorithm to frequency domain by adjusting the transmit frequency. Rahmati et al.~\cite{rahmati2007context} selectively turn on wireless radios based on the probability distribution of Wi-Fi connectivity, which is estimated based on user's context information. Likewise, Nicholson et al~\cite{nicholson2008breadcrumbs} predicts Wi-Fi connectivity based on user's previous mobility traces.
HQS in~\cite{hqs2014} extends Quorum-based power-saving (QPS) for 802.11 ad hoc mode to arbitrary cycle lengths to increase energy efficiency.

\textbf{Downclocking.} Recently, researchers have demonstrated that downclocking Wi-Fi radios can effectively reduce the energy consumption during packet reception and idle listening. E-Mili~\cite{mobicom11emili} pioneers this kind of mechanisms to downclock receiver's clock rate during idle listening, and switches to full clock rate for packet reception. SloMo~\cite{nsdi13slomo} enables packet decoding in IEEE 802.11b while downclocking by exploiting the sparsity in direct sequence spread spectrum (DSSS) PHY. SEER~\cite{wang2017wideband} designs a special preamble to allow narrowband receiver sensing a much wider band signal without boosting the clock rate. Downclocking for packet reception has been extended to OFDM-based Wi-Fi by leveraging the gap between modulation and SNR~\cite{mobicom14enfold}. Instead of exploiting PHY redundancy, Sampleless Wi-Fi~\cite{wang2017sampleless,wang2016rateless} utilizes the retransmission opportunities to decode packets at downclocked rates without relying on PHY redundancy, and thus makes downclocking OFDM packet reception feasible under low SNR conditions. \texttt{T-Fi} takes an opposite approach by overclocking the receiver's radio to reduce the power consumption of the transmitter, which is complementary to downclocking approaches by bringing low power to the transmitter side.

\vspace{0.3cm}

\section{\textbf{Discussion}}
\label{sec:discussion}

\textbf{The benefit of \texttt{T-Fi} to IoT devices.} Lots of OFDM-based IoT devices can be benefited from \texttt{T-Fi} design. In particular, there are a large number of commercial IoT devices supporting Wi-Fi protocols, including smartwatches~\cite{Apple_watch}, wireless cameras~\cite{arlo}, smart loudspeakers~\cite{echo_dot}, and virtual-reality (VR) headsets~\cite{Magic}. These devices implement Wi-Fi to upload large amount of data. \texttt{T-Fi} can prolong the lifetime of these devices. Besides, after IEEE 802.11ah is deployed in low-power IoT devices, they can also be benefited from \texttt{T-Fi}. Finally, the asymmetry between IoT devices and gateways is ubiquitous. Therefore, other protocols can leverage oversampling to save transmission power of IoT devices, with tailored decoding pipelines.

\textbf{Power gain.} In theory, doubling the clock rate yields power gain of 3~dB, which, however, does not hold in our experiments due to two reasons. First, the overall performance gain of the whole system is related to all algorithms in the reception pipeline. The performance of some algorithms other than the decoding algorithm depends on the SNR. Second, the theoretical analysis only considers the influence of Gaussian noise and supposes the noise is independent in time domain, which does not hold in practice due to the hardware defect~\cite{renani2017harnessing} and environmental interference~\cite{khan2016accurate}.

\textbf{Multi-antenna and high-sensitivity receivers.} Multiple antennas and high-sensitivity receivers can also be used to improve the decoding performance and save the transmission power. However, these solutions require additional hardware to be deployed on the AP side, while our target is to makes full use of the inherent redundant resource readily available in commercial APs. In addition, \texttt{T-Fi} is orthogonal with these solutions and can be easily integrated with these solutions to deliver extra performance gain.

\vspace{0.3cm}

\section{Conclusion}\label{sec:conclusion}
This paper introduces \texttt{T-Fi}, an asymmetric transceiver paradigm for IoT that pushes power burden to the AP side, and enables IoT devices to transmit packets at power levels that are even lower than the minimal power required by conventional receivers. We think this is an important design point in IoT communications, and the \textbf{asymmetric} fashion in \texttt{T-Fi} has significant ramifications in the new transceiver design for IoT Wi-Fi protocols. Our experimental evaluation confirms the benefits of \texttt{T-Fi} in real environments. We hope the design can contribute the wireless community by providing some insights for future transceiver design that takes advantage of the hardware asymmetry between APs and IoT devices.

\balance
\bibliographystyle{IEEEtran}
\bibliography{IEEEabrv,./oversample}

\begin{IEEEbiography}[{\includegraphics[width=1in,height=1.25in,clip,keepaspectratio]{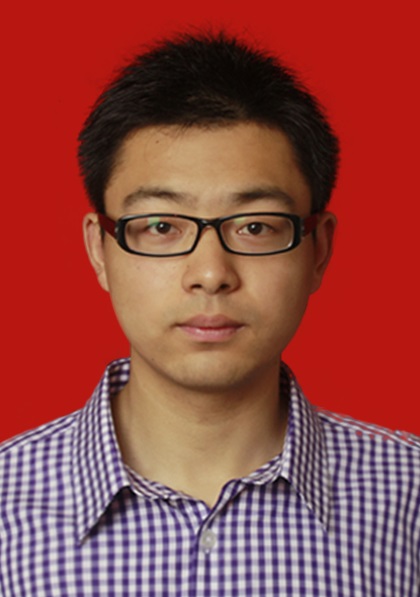}}]
	{Wei Wang (S'10-M'16)} received the Ph.D. degree from the Department of Computer Science and Engineering, The Hong Kong University of Science and Technology. He is currently a Professor with the School of Electronic Information and Communications, Huazhong University of Science and Technology. His research interests include PHY/MAC design and mobile computing in wireless systems. He served on TPC of INFOCOM and GBLOBECOM. He served as Editors for IJCS, China Communications, and Guest Editors for Wireless Communications and Mobile Computing and the IEEE COMSOC MMTC COMMUNICATIONS.
\end{IEEEbiography}

\begin{IEEEbiography}
	[{\includegraphics[width=1in,height=1.25in,clip,keepaspectratio]{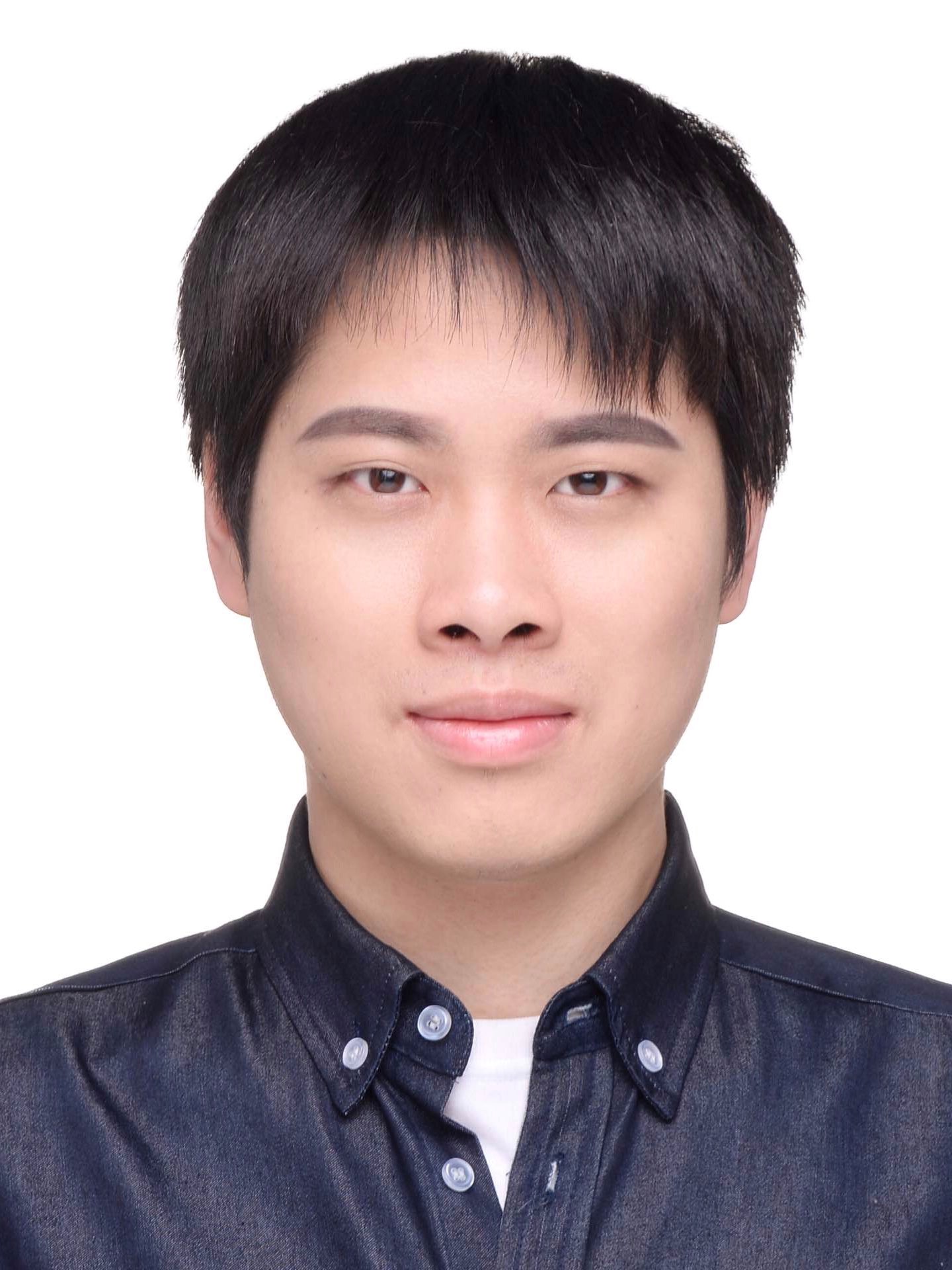}}]
	{Shiyue He} is currently pursuing his Ph.D. degree at School of Electronics and Information Engineering, Huazhong University of Science and Technology, Hubei, China. Before that, he has received his Bachelors degree in information engineering from Wuhan University of Technology, Hubei, China, in June 2016. His research interests include UAV communications and beamforming in wireless networks.
\end{IEEEbiography}

\begin{IEEEbiography}[{\includegraphics[width=1in,height=1.25in,clip,keepaspectratio]{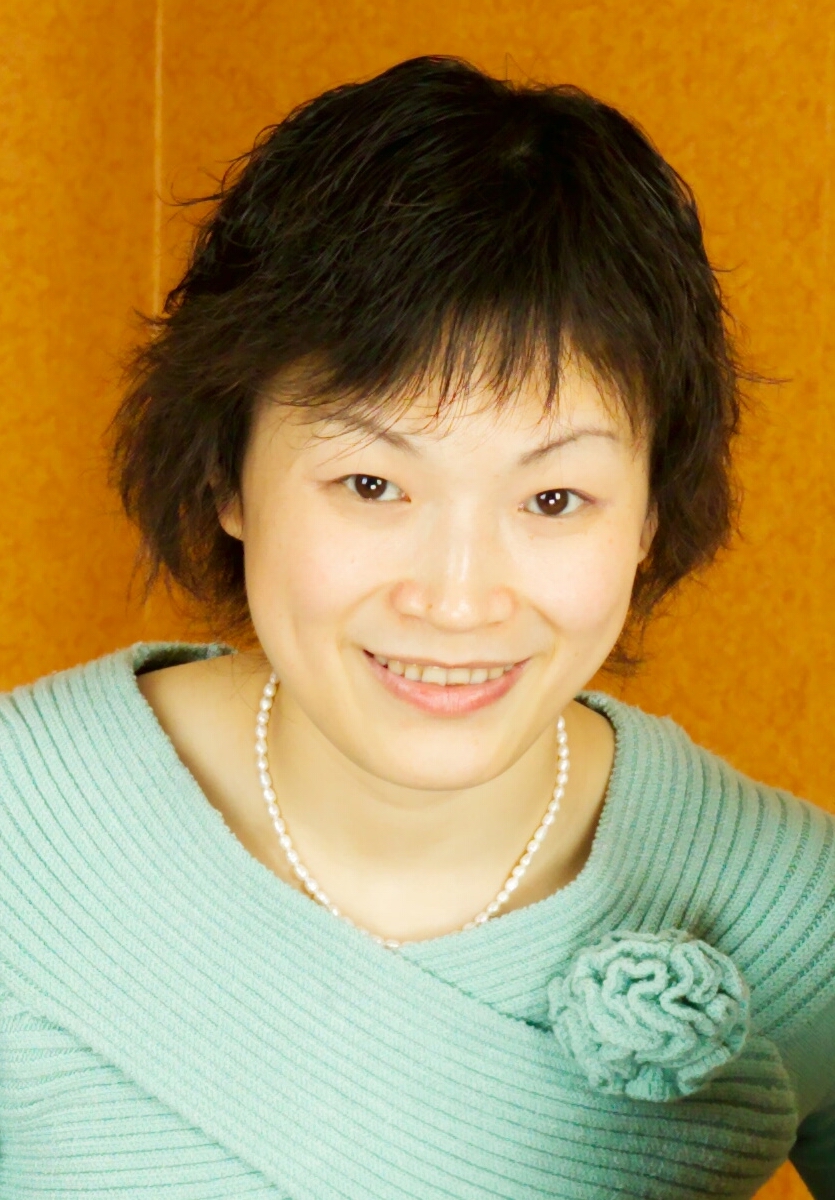}}]
	{Qian Zhang (M'00-SM'04-F'12)} joined Hong Kong University of Science and Technology in Sept. 2005 where she is a full Professor in the Department of Computer Science and Engineering. Before that, she was in Microsoft Research Asia, Beijing, from July 1999, where she was the research manager of the Wireless and Networking Group. She is a Fellow of IEEE for ``contribution to the mobility and spectrum management of wireless networks and mobile communications''. Dr. Zhang received the B.S., M.S., and Ph.D. degrees from Wuhan University, China, in 1994, 1996, and 1999, respectively, all in computer science.
\end{IEEEbiography}

\begin{IEEEbiography}[{\includegraphics[width=1in,height=1.25in,clip,keepaspectratio]{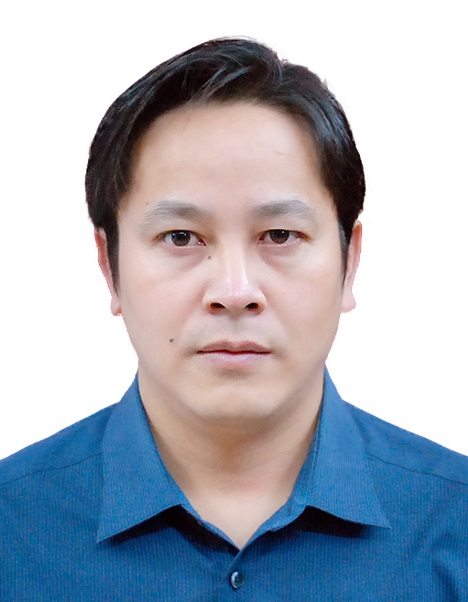}}]
	{Tao Jiang (M'06-SM'10-F'19)} is currently a Distinguished Professor in the Wuhan National Laboratory for Optoelectronics and School of Electronics Information and Communications, Huazhong University of Science and Technology. He received Ph.D. degree in information and communication engineering from Huazhong University of Science and Technology in April 2004. From Aug. 2004 to Dec. 2007, he worked in Brunel University and University of Michigan-Dearborn, respectively. He served as associate editors in IEEE Transactions on Signal Processing and IEEE Communications Surveys and Tutorials. He is a recipient of the NSFC for Distinguished Young Scholars Award in 2013, and he is a Fellow of IEEE.
\end{IEEEbiography}

\end{document}